\documentclass[mathleft, fleqn, final
]{an}
\usepackage{graphicx}
\usepackage{times}
\usepackage{amsmath}
\usepackage{ulem}
\usepackage{natbib}
\usepackage{subfigure}
\def \carmenes {CARMENES}
\def \mps {$\mathrm{ms}^{-1}$}

\def \atlaswidth {15cm}

\begin{document}

\Pagespan{789}{}
\Yearpublication{2006}%
\Yearsubmission{2005}%
\Month{11}%
\Volume{999}%
\Issue{88}%

\title{On the optimum operating conditions of ThNe calibration lamps for measurements of radial velocity variations}

\author{M. Ammler-von Eiff\inst{1,2}\fnmsep\thanks{Corresponding author:
  \email{ammler@mps.mpg.de}\newline}
\and  E.W. Guenther\inst{1,3}
}
\titlerunning{Wavelength calibration with ThNe lamps}
\authorrunning{Ammler-von Eiff et~al.}
   \institute{Th\"uringer Landessternwarte, Sternwarte 5, 07778 Tautenburg, Germany
\and
Max-Planck-Institut f\"ur Sonnensystemforschung, Justus-von-Liebig-Weg 3, 37077 G\"ottingen, Germany\\
\email{ammler@mps.mpg.de}
\and
\"Osterreichische Akademie der Wissenschaften, Institut f\"ur Weltraumforschung, IWF, Schmiedlstra{\ss}e 6, 8042 Graz, Austria
         }

\received{30 May 2005}
\accepted{11 Nov 2005}
\publonline{later}


\abstract{
 Previous studies have shown that ThNe lamps are very suitable for the precise measurement of wavelength variations.
 A critical issue for all hollow cathode lamps (HCL) is the current at which the lamp is operated -- a low value has the advantage that the life time of the lamp is longer whereas the number of useful lines is lower. We investigate the number of suitable spectral lines depending on the current used and obtained spectra of a ThNe HCL coupled by a fibre to the Tautenburg Echelle spectrograph in a setting that is typical for many high-resolution spectrographs. Operating currents were chosen in the range specified by the manufacturer. We varied exposure time to identify the saturation level and effects of noise.
A few thousands of Th lines were identified automatically in the wavelength range considered. We noticed a scatter of several hundred ms$^{-1}$ of Th lines when varying the settings but did not detect any systematic trends. The scatter in wavelength residuals of Th lines however indicates that a precise control of operating current can be necessary. An estimation of the calibration precision of individual lines indicates that a combination of strong Th lines allows one to reach a level of 1\,\mps. Although a high operating current reduces the life-time of the lamp and accelerates its ageing it guarantees highest precision because of the numerous Th lines which become strong compared to the Ne lines. Then, noise, saturation, wavelength residuals of lines identified, and the pollution due to saturated Ne lines can be minimised by adjusting exposure time. Our results indicate that ThNe lamps can be better suited than ThAr lamps for all kinds of studies that involve the precise measurement of radial velocity variations, e.g. studies on late-type stars, brown dwarfs, objects that suffer from strong extinction, or high-redshift galaxies.
}

\maketitle

\section{Introduction}

We characterise the spectrum of a ThNe hollow cathode lamp (HCL) which is an interesting choice for new spectrographs extending beyond the red part of the spectrum into the $Y$ band. One of the main science drivers to use this wavelength range is the search for planets in the habitable zone around M dwarfs which requires a long-term precision of $1\,\mathrm{ms}^{-1}$ in radial velocity measurements. Examples are given by HPF\footnote{\underline{H}abitable-zone \underline{P}lanet \underline{F}inder}, a near-infrared (NIR) spectrograph for the Hobby-Eberly-Telescope \citep{2012SPIE.8446E..1SM}, {\carmenes}\footnote{\underline{C}alar \underline{A}lto high-\underline{R}esolution search for \underline{M} dwarfs with \underline{E}xo-earths with \underline{N}ear-infrared and optical \underline{E}chelle \underline{S}pectrographs} \citep{Quirrenbach+2012,2013hsa7.conf..842A}, a spectrograph for the 3.5m telescope at Calar Alto consisting of two arms for the visual and the NIR wavelength range, and SPIRou\footnote{\underline{S}pectro\underline{P}olarim\`etre \underline{I}nfra-\underline{Rou}ge} \citep{2011ASPC..448..771A}, a NIR spectro-polarimeter for the CFHT\footnote{\underline{C}anada \underline{F}rance \underline{H}awaii \underline{T}elescope}. All of those are fibre-fed and stabilised. A similar project which is already operational, is done with CYCLOPS\footnote{http://www.phys.unsw.edu.au/$\sim$cgt/CYCLOPS/CYCLOPS.html}, a fibre-feed for the UCLES\footnote{\underline{U}niversity \underline{C}ollege \underline{L}ondon \underline{E}chelle \underline{S}pectrograph} Echelle spectrograph at the Anglo-Australian Telescope. It covers a wavelength range of 540-785\,nm at a resolving power of 70,000.

The spectrum of a hollow cathode lamp (HCL) was first studied by \citet{Paschen1916} who identified the spectrum of He in the HCL discharge. \citet{Schueler1926} used the HCL discharge to produce a spectrum of metal vapour paving the way to the precise wavelength calibration of high-resolution spectrographs. \citet{2007ASPC..364..461K} reviewed the use of HCL for wavelength calibration and gave criteria for calibration sources. Thorium has been a common choice for the calibration of high-resolution spectrographs since it provides a dense forest of sharp spectral lines over a wide wavelength range. The wavelengths are well-known from laboratory measurements. 

In the past, much work has been devoted to the derivation of Th line lists. \citet{Palmer+1983} presented a catalogue of Th lines in the wavelength range $278-1350$\,nm obtained from FTS measurements of a ThNe HCL. Only the Th lines are used for wavelength calibration. They are sharp because of their high atomic weight. Neon has lower atomic weight so that the Ne lines are intrinsically wider under identical conditions \citep{Palmer+1983}. Therefore, Ne lines are less suitable for precise wavelength calibration than the Th lines. In the past decade, much effort has been devoted to improvements of the Th line list \citep{2007A&A...468.1115L,Redman+2014}. Moreover, the Th line list has been extended to the near-infrared \citep{Hinkle+2001,EnglemanJr20031,2007ASPC..364..461K,Kerber+2008} in consideration of the development of near-infrared high-resolution spectrographs.

Fourier transform spectroscopy (FTS), which is carried out at very high resolving power of the order of 1,000,000 and better, shows that there are blends which will be unresolved at a resolution of less than 100,000 typical of upcoming extended red spectrographs. Such blends are a severe complication to wavelength calibration \citep[cf.~][figure~6]{Redman+2014}. The relative intensities of blending lines may change with time (ageing) so that the line centre of the blend effectively changes.

The filling gas plays an important role since in practice, strong noble gas lines can deteriorate the wavelength calibration by affecting Th lines in the immediate neighbourhood or even blocking a whole spectral range. Although ThAr lamps are the traditional choice for wavelength calibration, the strength of the Ar lines poses a challenge to extended red spectrographs since the red part of the visual spectrum is polluted by numerous strong and saturated Ar lines. As shown in \citet{Mayor+2009} and \citet{2010aepr.confE..16L} for ThAr lamps, the ageing effects of lamps are more prominent for the argon lines than for the thorium lines.

The effects of the filling gas can be mitigated by another choice. In principle, any noble gas or a mixture is possible \citep{2007ASPC..364..461K}. The AAT search for M-dwarf planets uses Xe as filling gas. The ThXe HCL is calibrated using a ThAr lamp and then used for simultaneous calibration of the science spectra. In the present work, we investigate the use of Ne which is a well-studied alternative to Ar. \citet{2011ApJS..195...24R} showed that UNe lamps - a promising choice for the NIR - display Ne lines weaker than the Ar lines in the UAr spectra. Also the KPNO HeNeAr atlas\footnote{http://old-www.noao.edu/kpno/specatlas/henear/henear.html} nicely illustrates that the Ne lines are not as strong as the Ar lines. 

The spectrum of HCLs depends strongly on operating current. High operating current maximises the number of available lines but the life time of the lamps and probably also the ageing scales with reciprocal current \citep{2007ASPC..364..461K}. For this reason, the choice of an operating current as low as possible is desired. Therefore, a detailed study of the ThNe spectrum is necessary and an optimisation of the operating current is required.
\citet{2007ASPC..364..461K} investigated how the spectrum of the ThAr HCL changes with current in the wavelength region from 900 to 4,500\,nm. They found that the average intensity of Th lines changes linearly with the operating current but the Ar lines change non-linearly. This means that the intensity ratio of the Th and the Ar lines depends strongly on the current used.

To our knowledge, this kind of analysis has not yet been done for ThNe lamps.  The goal of the present work is to identify a value of the operating current which maximises the number of useful spectral lines and at the same time does not affect too much the life time of the lamp.

It is desired to use the lamps at the minimum current that is still useful. The question is whether it is better to use a lower current, and thus a longer exposure time, or to use a higher current and thus a shorter exposure time. Because the spectrum changes with current, it is crucial to obtain spectra at different current in order to select the best operation mode.

We take into account the pollution by Ne lines and identify wavelength ranges which should be excluded from wavelength calibration because of the presence of individual strong Ne lines. Furthermore, we account for the effects of noise and saturation, and try to minimise those. The background level is corrected for but not of interest for wavelength calibration and line identification. Therefore, it is not discussed in the present work (but see the study of \citealp{2007ASPC..364..461K}). 

We not only study the number and strength of lines but also show how exposure time and operating current of the calibration lamp affect the attainable RV precision.
\citet{Butler+1996} and \citet{Bouchy+2001} show how the line strength and the noise level impact the achievable RV precision when measuring the relative shifts of lines in stellar spectra. The achievable precision of RV measurements is determined by the ability to measure intensity variations obtained at every pixel of the spectrum. The sensitivity scales with the intensity change and the noise level of the spectrum. Hence, the steepest parts of spectral lines and thus the strongest lines are most sensitive to RV variations. Of course, the overall RV precision achievable should be limited by the stellar spectrum and not by the calibration spectrum. Therefore, when using Th lines, their positions have to be determined very precisely -- not absolutely but in a relative sense. As for the stellar lines, this works best when the lines are sufficiently strong. This allows us to apply the recipes given by \citet{Butler+1996} and \citet{Bouchy+2001} to investigate how the choice of operating current impacts the attainable RV precision.


The paper is organised in the following way. Sect.~\ref{sect:measure} explains the spectroscopic measurements, the data reduction, the line identification, the line measurements, and presents an atlas of Th lines. Then, the density of lines in different parts of the spectrum is studied (Sect.~\ref{sect:lines}) and the regimes of noise and saturation are assessed (Sect.~\ref{sect:sat}). The residuals of wavelengths of identified lines (Sect.~\ref{sect:wave}), RV precision (Sect.~\ref{sect:RV}), number, and the distribution of line strength (Sect.~\ref{sect:strengths}) are analysed under a variation of exposure time and operating current before we conclude in Sect.~\ref{sect:concl}.

\section{Spectroscopic measurements, data reduction, and line identification}
\label{sect:measure}

Spectra were obtained using the calibration unit built for {\carmenes} \citep{Quirrenbach+2012,2013hsa7.conf..842A}. This calibration unit injects light from an HCL into a 100\,$\mu$m fibre which we connected to the standard Echelle spectrograph of the 2m telescope in Tautenburg. Therefore, the setup is a model for {\carmenes} and all fibre-fed spectrographs with a similar design of the calibration unit. A number of spectra of a ThNe HCL were taken this way in Nov 2012 at currents of 3, 6, 9, and 12\,mA and exposure times of 1, 10, and 60\,s. The spectrograph operates at visual wavelengths but offers a red channel which extends the spectral range towards the NIR ($5,440-10,250$\,{\AA}). The resulting resolution is slightly higher than for standard stellar spectroscopy at the 2m telescope in Tautenburg. Using a few Th lines, we measured a resolving power of 65,000 - 75,000 which is similar to that of upcoming extended red spectrographs. For the line identification and all calculations below, a value of 70,000 has been adopted. The temperature of the spectrograph is stabilised and varies by about 0.1\,K at most. All spectra were bias-subtracted and the Echelle orders extracted using standard routines in IRAF\footnote{IRAF is distributed by the National Optical Astronomy Observatories, which are operated by the Association of Universities for Research in Astronomy, Inc., under cooperative agreement with the National Science Foundation.} \citep{iraf86,Tody1993}. The spectrum for 12\,mA and 10\,s was used to derive the wavelength solution which was applied to all the ThNe spectra. This is sufficient for the purpose of the present work since we are not interested in a very precise absolute calibration but only in the relative variation of wavelengths when changing operating current.

\begin{table}
\caption{\label{tab:list}
The table summarizes the contents of the line list in the wavelength range studied: {\bf (1)} species; {\bf (2)} number of lines; {\bf (3)} source of line data. The table includes the 63 strong Ne lines excluded later on.}
\centering
\begin{tabular}{lrl}
\hline
species & \# & source\\
\hline
Ne\,I   &   89 & NIST \citep{Saloman+2004}\\
Ne\,I   &    3 & \citet[][table~5]{Redman+2014}\\
Ne\,II  &   14 & NIST \citep{Persson1971}\\
Ne\,II  &    4 & \citet[][table~5]{Redman+2014}\\
Th\,I   &3,754 & \citet[][table~6]{Redman+2014}\\
Th\,II  &  981 & \citet[][table~6]{Redman+2014}\\
Th\,III &   41 & \citet[][table~6]{Redman+2014}\\
\hline
\end{tabular}
\end{table}

The identification and measurement of spectral lines was done with IRAF recipes and own tools written in IDL as is described below. In each one of the ThNe spectra, thorium and neon lines were identified from tables 5 (neon) and 6 (thorium) of \citet{Redman+2014}\footnote{Before we learned about the work of \citet{Redman+2014}, we identified Th lines from a digital list of Th I-III lines kindly provided by R. Engleman (priv. comm.), a list which is based on the \citep{Palmer+1983} atlas of Th lines covering a wavelength range of $2,777-13,500$\,{\AA} and further includes many lines of Th III from \citet{Behrens+1989}.}. Furthermore, we included a list of 390 Ne\,I and Ne\,II lines from the NIST database \citep{Persson1971,Saloman+2004}\footnote{http://physics.nist.gov/PhysRefData/Handbook/Tables/neontable2\_a.htm}.

\citet{Redman+2014} recommended the use of Ritz wavelengths for line identification and wavelength calibration. However, Ritz wavelengths are available only for a subset of the lines in our list. Furthermore, we are not interested in a high absolute precision of wavelength calibration, since we only study the (relative) variation of wavelengths. Therefore, we decided to use observed wavenumbers. Since our measurements were taken at atmospheric pressure, the wavenumbers were converted to air wavelengths using the formula of \citet{EDLEN:53} and the coefficients of \citet{Ciddor:96}. From the combined ThNe line list, we removed duplicate lines and flagged blends given the resolving power of the spectrograph (70,000). This leaves a total of 4,776 isolated Th lines and 369 blends in the wavelength range of the spectrograph. Table~\ref{tab:list} lists the total numbers of lines sorted by chemical species.


\begin{figure}
\includegraphics[width=\hsize]{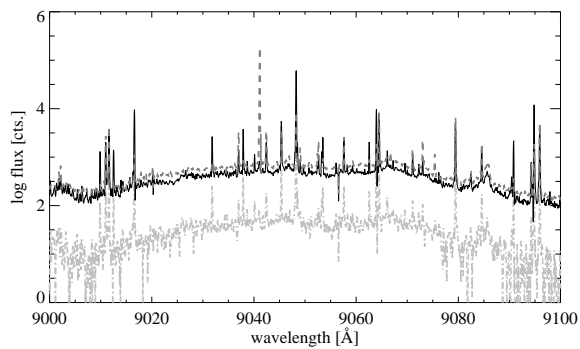}
\caption{\label{fig:plot_spec_sample} A chunk of the ThNe spectrum in the red is shown for different exposure times of 1\,s (grey line) and 10\,s (solid line) at an operating current of 12\,mA, and for 60\,s at an operating current of 3\,mA (dashed line). Note that the short exposure taken at an operating current of 12\,mA is affected substantially by noise. The background flux level reaches a few tens of counts in the short exposure and a few hundreds of counts in the long exposures.}
\end{figure}

We identified lines not via line strength but exclusively based on wavelength since the line lists from literature were obtained at other values of operating current. Note that \citet{Redman+2014} used a current of 25\,mA whereas we used 12\,mA at maximum. This is because the maximum current recommended by the manufacturer is 15\,mA for the lamp that we used. We thus hesitated to apply any higher current or to operate it at the maximum current. It is known from previous work and we will see in Sect.~\ref{sect:results} that the relative strength of lines strongly depends on operating current (Fig.~\ref{fig:plot_spec_sample}).

The identification of spectral lines was done fully automatically with the IRAF task onedspec.identify. The spectrum was searched for lines close to the wavelength positions given in the line list\footnote{using the interactive key command \textit{e} of the IRAF onedspec.identify recipe and the command \textit{features} of the task to obtain the central wavelengths of spectral lines identified.}. The task measures the central positions of the lines identified which can deviate slightly from the positions given in the literature.

The number of features identified varies between 2,100 and 3,100 depending on operating current and exposure time, found close to the laboratory wavelengths with an rms of $0.05-0.07\,${\AA}.

An atlas of Th lines is shown in Appendix~\ref{app:A}. It is a ThNe spectrum similar to the one expected for {\carmenes} and similar spectrographs in terms of wavelength range, resolution, and relative line strengths. We chose the spectrum obtained at an operating current of 12\,mA and an exposure time of 10\,s. As we will see later on, this is a setting found to provide a large number of Th lines useful for wavelength calibration. In addition, we highlighted spectral windows polluted by very strong Ne lines. The saturating Ne lines affect whole columns on the CCD and thus other spectral orders. Those signatures have been removed manually in the atlas and have been excluded from the analyses described below. The atlas shows chunks of 50\,{\AA} and gives the number of lines in these chunks. For clarity, the plotted spectra overlap by 10\,{\AA} and only lines with a valid measurement of line strength (see below) were included. Labels were placed at the wavelengths measured in the identification process. As those usually do not fully coincide with the laboratory wavelengths, the amount of deviation is coded by line style (dashed: less than half of the mean residual of $0.07\,${\AA} dot-dashed: less than the mean residual; dotted: more than the mean residual). The dotted horizontal line indicates the approximate saturation limit of 5.4\,dex in logarithmic scale. 

There are 63 grey-shaded regions dominated by strong Ne lines with damping wings. They have been identified by visual inspection. The measured wavelengths of Th lines in the wings of these Ne lines will be spurious. Therefore, regions with widths of 2, 4, or 8\,{\AA}, resp., were excluded around each of those Ne lines, depending on the width of the Ne line. Hence, 349 lines in those regions were excluded from the studies below.

Then, line strengths were obtained by adopting the peak values (minus the background level) instead of measuring the total flux within a line. This allows us to identify saturating lines and to account for saturation in the following steps. For the Th lines at moderate strength, which we are interested in, the peak value still scales linearly with the flux. The peak values were obtained via Gaussian profiles matching the emission lines and accounting for the background level. The instrumental profile of the spectrograph is not perfectly symmetric so that wavelength measurements may be affected by a small constant offset. This is acceptable since we are only interested in the relative shift of wavelengths and not in their absolute values.

\section{Results and Discussion}
\label{sect:results}
\subsection{Line density}
\label{sect:lines}


\begin{figure}
\includegraphics[width=\hsize]{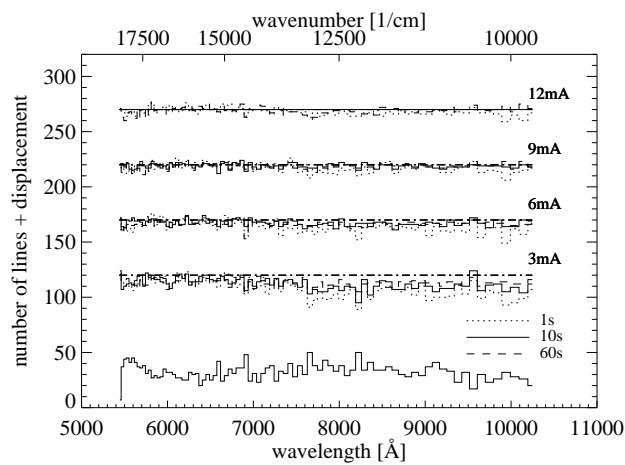}
\caption{\label{fig:density} Distribution of Th lines for different combinations of exposure time and operating current as indicated in the plot. The full distribution is only shown for an operating current of 12\,mA and an exposure time of 10\,s. It serves as reference for all the other combinations for which the residuals are shown only. Exposure time is coded by line style as indicated in the legend. For clarity, the residuals are displaced by 120 plus multiples of 50. The bin size is 100\,cm$^{-1}$ in wavenumber units.}
\end{figure}

The moderate overall increase of the number of Th lines identified with operating current and exposure time is reflected in the line density (Fig.~\ref{fig:density}). The density hardly changes from the blue to the red except for the low currents which are susceptible to variations of the sensitivity of the spectrograph (see distribution of line strength in Fig.~\ref{fig:stat_strength2}). The sensitivity peaks close to 7,000\,{\AA} and decreases towards redder wavelengths. 

The distribution is shown in constant bins of wavenumber to enable a direct comparison with \citet[][fig.~6]{2011ApJS..195...24R} who show the data of \citet{Kerber+2008}. Note however that those data were obtained in another setting which is more sensitive in the red.

\subsection{Saturation and noise}
\label{sect:sat}
The regimes of noise and saturation are explored to study the influence of exposure time, and to ultimately select Th lines useful for wavelength calibration that can be studied under changes of operating current. The comparison of line strengths measured at different exposure time reveals that saturation starts at about 5.4 in logarithmic scale, corresponding to some 100,000's of counts (digital counts or analogue-to-digital units). Saturation prevails among the Ne\,I lines. At an exposure time of 60\,s, many Th lines enter the saturation regime, too. Of course, the exact values are valid only for the Tautenburg setup but the general conclusions apply to other spectrographs as well.

As is the case for the saturated lines, the measured strengths of weak lines (less than $\sim$100 counts) do not scale according to exposure time. Then, line strengths are subject to noise or noise features are mistaken for spectral lines. Considering the gain of $2.6\,\mathrm{e}^{-}\mathrm{DN}^{-1}$ and the read-out noise of $4.16\,\mathrm{e}^{-}$ of the detector in Tautenburg, the limit of 100 counts corresponds to a signal-to-noise ratio of 15 (per pixel). It can be noticed that noise is an issue close to the blue (below 6,000\,{\AA}) and red cut-offs (above 9,500\,{\AA}) of the wavelength range. The wavelengths of these lines will be erroneous, so that these regions should be excluded from a wavelength calibration.

\begin{figure}
\subfigure{\includegraphics[width=\hsize]{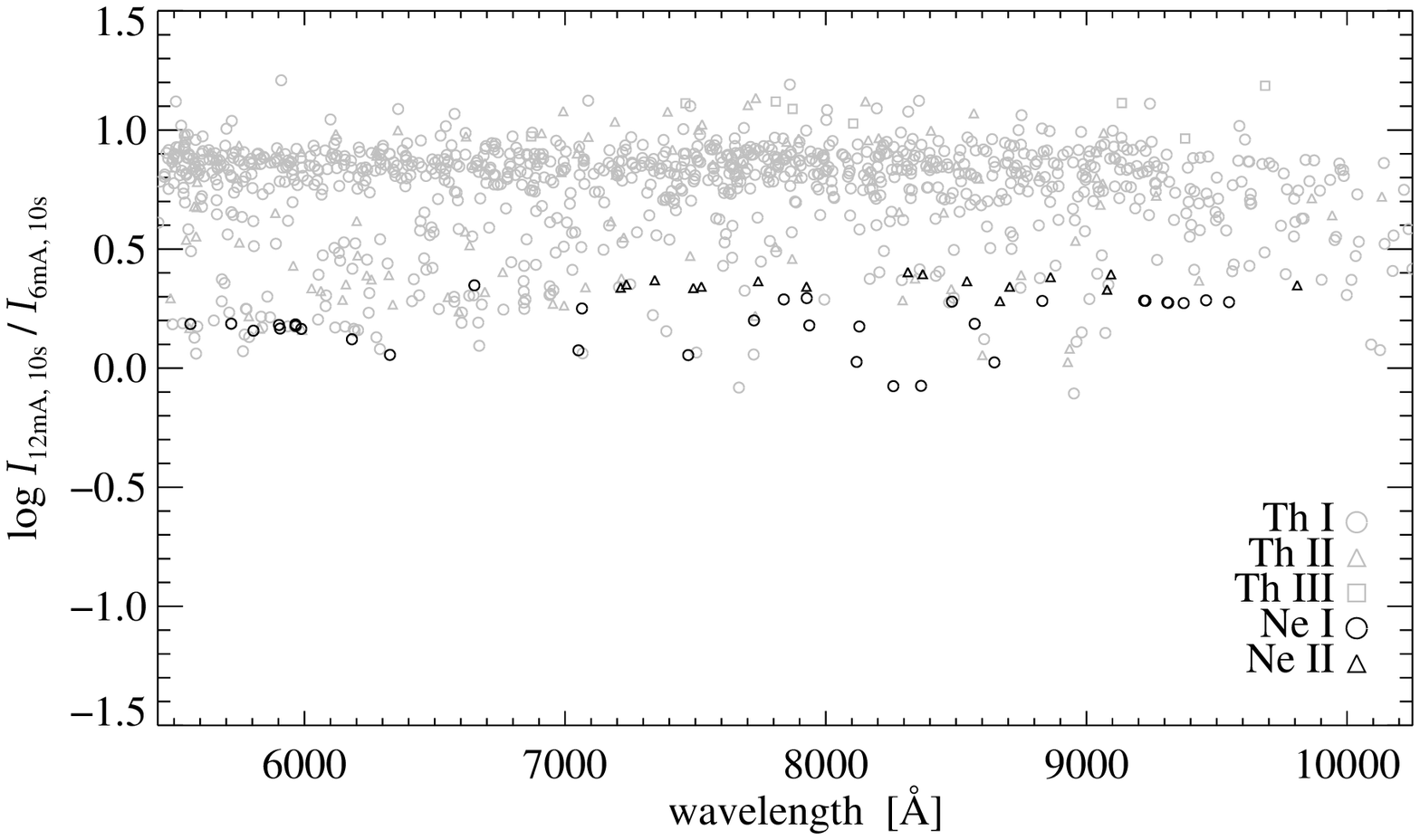}}
\subfigure{\includegraphics[width=\hsize]{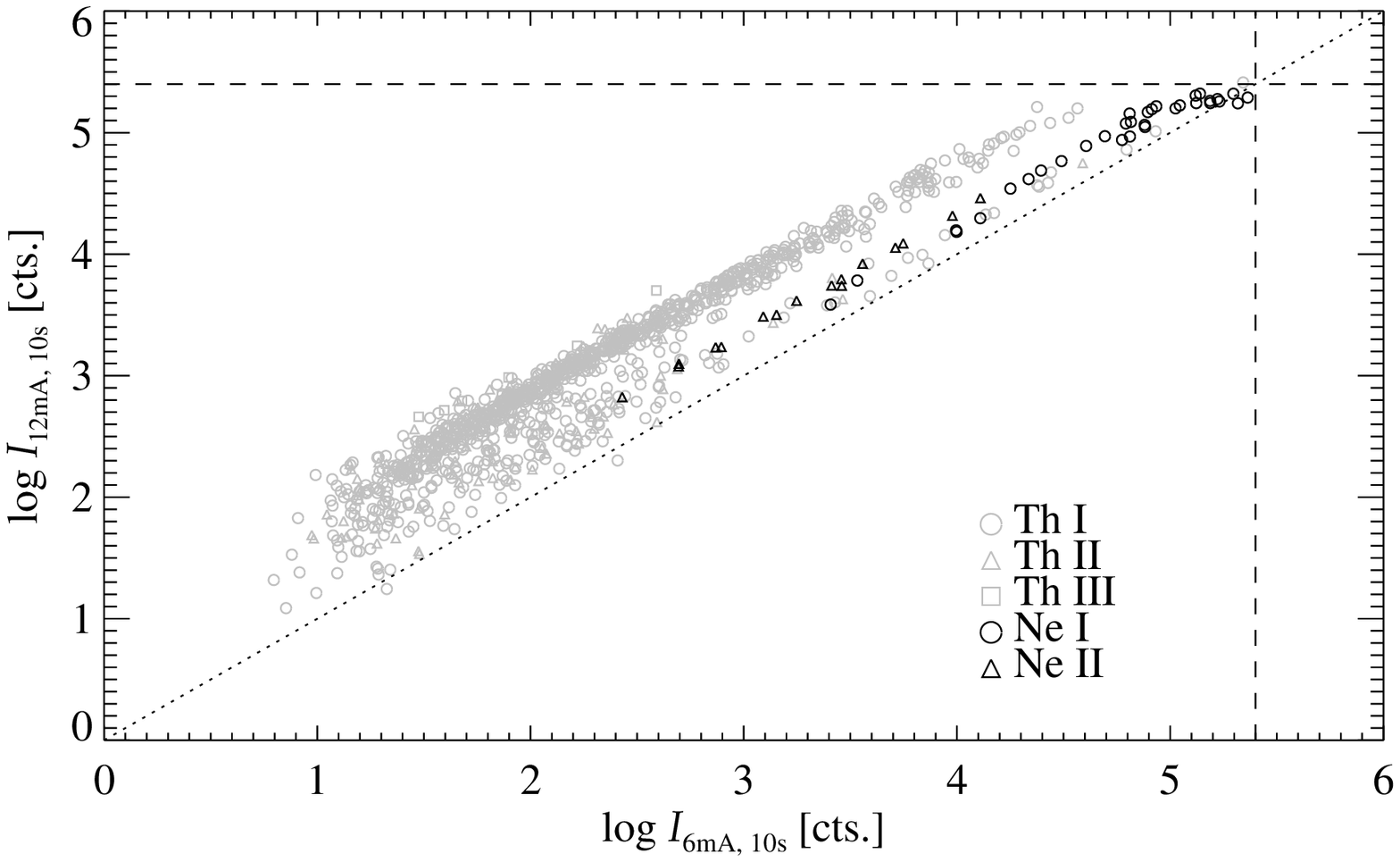}}
\caption{\label{fig:12_6} Comparison of two ThNe spectra obtained at different operating currents of 12\,mA and 6\,mA but with the same exposure time of 10\,s. The top panel displays the log. ratio of the strength of lines present in both spectra. Atomic and ionic species of Th and Ne are indicated by symbols. The bottom panel compares the log. line strengths directly. The dashed lines highlight the saturation limit and the dotted line indicates identity.}
\end{figure}

We compare ThNe spectra taken at different operating current in Fig.~\ref{fig:12_6}. The ratio of line strength in the top panel shows some structure that can be understood when consulting the bottom panel. A large group of Ne\,I lines is strong, mostly saturated in both spectra. The other Ne\,I lines as well as the Ne\,II lines are stronger by about 0.5\,dex at higher operating current. While a group of Th lines scales in a similar way, a large number of Th lines gets even stronger, by almost a factor of 10 on average, regardless of the ionisation stage. This reflects in a qualitative way that the average strength of Th lines increases more strongly with operating current than the average strength of the gas lines \citep[][here neon]{2007ASPC..364..461K}.

\subsection{Wavelengths}
\label{sect:wave}
\begin{figure}
\includegraphics[width=\hsize]{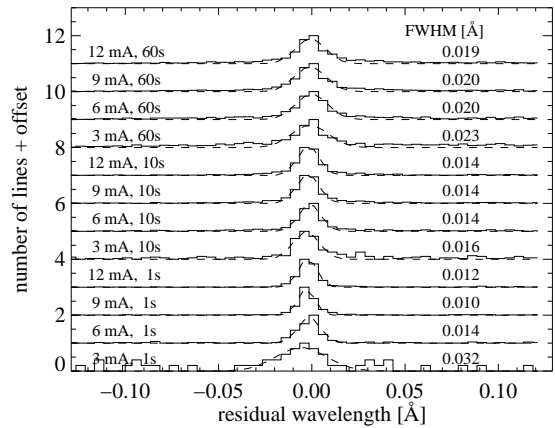}
\caption{\label{fig:wres} Distribution of wavelength residuals of Th lines identified. The histograms (solid lines) show the distribution for each spectrum, taken for different settings as indicated, and shifted by arbitrary offsets for clarity. The histograms are normalised to a peak value of 1 since the number of lines increases strongly with operating current. The full width at half maximum (FWHM) is estimated from best-matching Gaussian curves (dashed lines).}
\end{figure}




In the following, the variation of wavelengths is considered under changes of operating current and exposure time. Such variations may occur in the case of unresolved blends. In particular, we are interested in the presence of Th lines that are composed of two or more components. Based on the noise and saturation limits assessed in Sect.~\ref{sect:sat}, only lines stronger than the noise limit (100\,counts) and fainter than the saturation limit (5.4\,dex in log scale) are included in the following. For each line identified, the deviation from the laboratory wavelength has been measured. For each spectrum, the residuals of Th lines pile up close to a deviation of zero, i.e. according to expectations, there is no obvious systematic trend with operating current or exposure time on average (Fig.~\ref{fig:wres}). In contrast to the value of $0.05-0.07$\,{\AA} found above (Sect.~\ref{sect:measure}), the scatter is much less ($\sim0.01-0.03\,${\AA}) since only Th lines are considered and noisy and saturated lines were excluded. The scatter is least for high operating currents but increases with exposure time. Nevertheless, this statistical information does not tell anything about the behaviour of individual lines when settings are changed.

To better assess the behaviour of individual lines, lines were selected which are measurable in all spectra, comprising 203 Th lines and 38 Ne lines. Here, we could not exclude noisy and saturating lines since most of the lines are noisy or saturate in one spectrum or another. Certainly, this is due to the high dynamic range implied by the different exposure times and operating currents (cf. Fig.~\ref{fig:stat_strength2}).

Under varying operating settings, we found for those lines that the distribution of wavelength shifts cannot be distinguished from random distributions. The standard deviation is of the order of 10\,m{\AA} (median of 8\,m{\AA} for Th and 11\,m{\AA} for Ne lines). Nevertheless, we cannot exclude with the data at hand that the distributions are formed by systematic shifts of spectral lines with operating current. Those cannot be studied with the spectra available since the sampling of exposure times (3 values) and operating currents (4 values) is too low.

The scatter at the level of 10\,m{\AA} is much higher than that found by \citet{Kerber+2008}. They measured wavelengths of Th lines with ThAr lamps operated at a current of 20\,mA. The difference between their values and those of \citet{EnglemanJr20031} is much less (at a level of $10^{-8}$ in a relative sense), despite another choice of fill gas, pressure, and operating current. Regarding this previous result, the outcome of the present work is not understood. However, we note that we obtained measurements with an Echelle spectrograph while the previous work is based on FTS spectra. Certainly, there will be an effect due to the inclusion of noisy and saturated lines in the present analysis. The discussion of Fig.~\ref{fig:wres} showed that the scatter reduces substantially when excluding noisy and saturated lines.

In terms of radial velocity studies, the absolute accuracy is not that important as long as the position of Th lines can be reproduced precisely from spectrum to spectrum \citep{Redman+2014}. Nevertheless, a variation of line positions under variable operating conditions can become important if operating current is not stable. Then, the wavelength residuals identified for Th lines would be of the order of several hundred \mps over the spectral range studied here.

\subsection{Calibration of radial velocity measurements}
\label{sect:RV}

\begin{figure*}
\subfigure{\includegraphics[width=8cm]{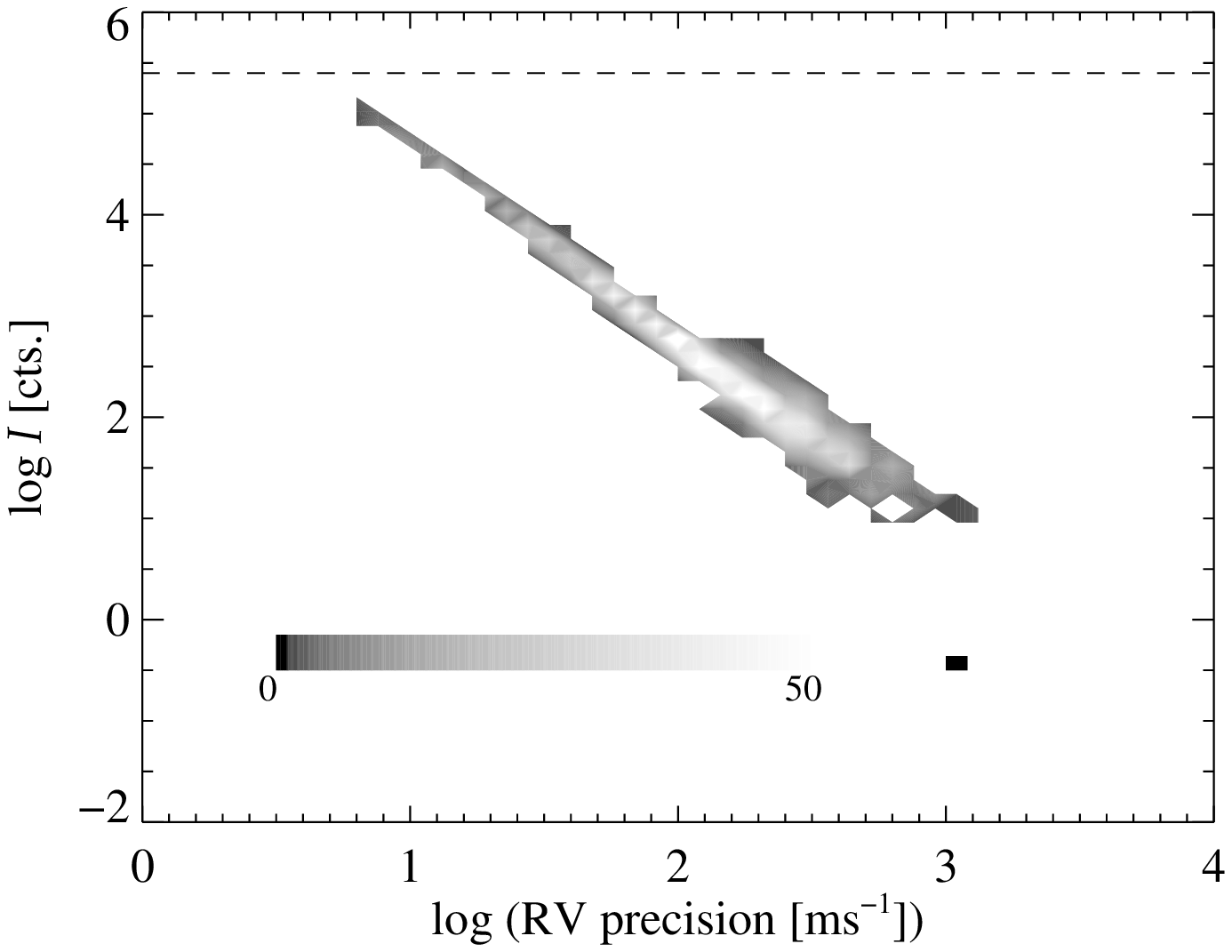}}
\subfigure{\includegraphics[width=8cm]{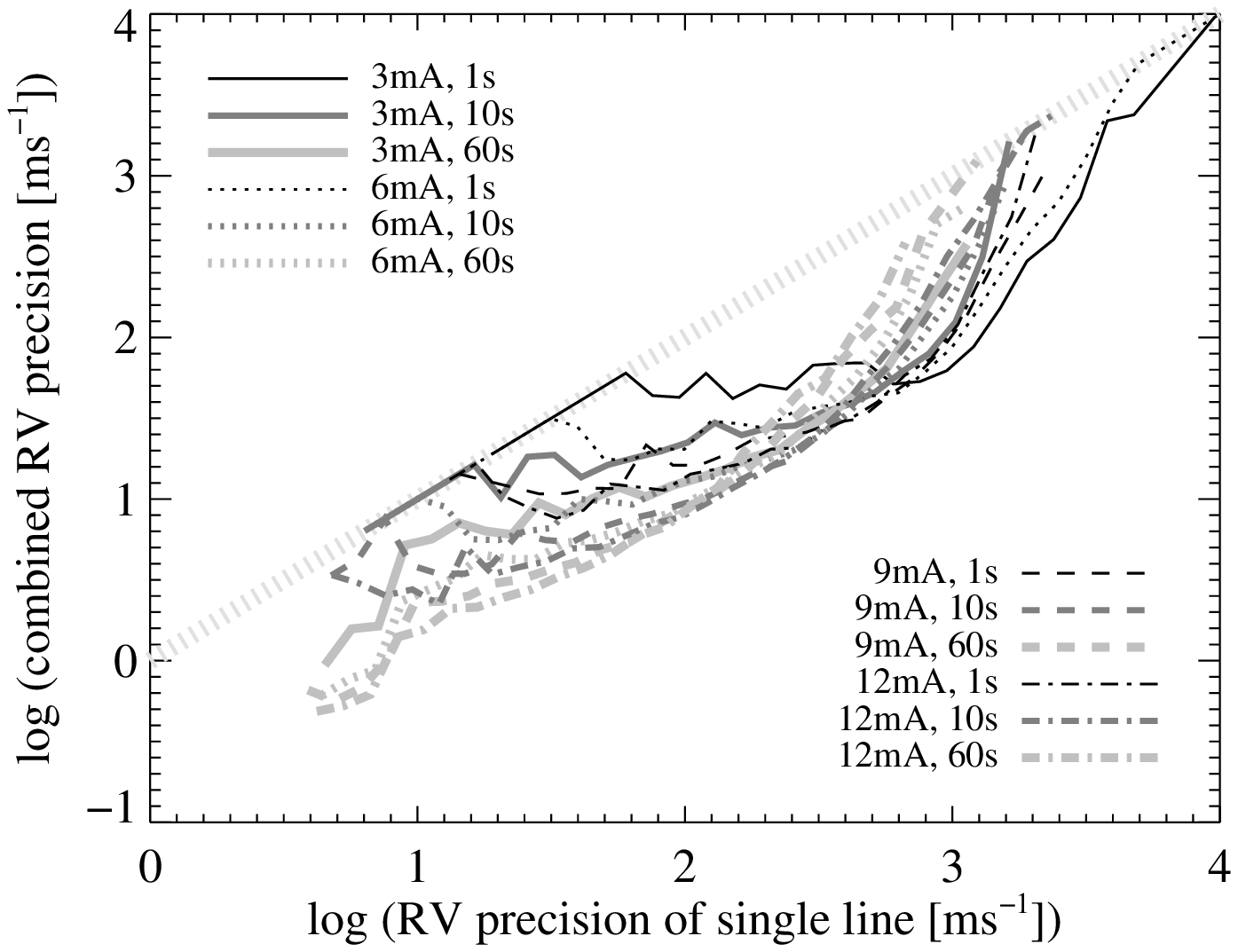}}
\caption{\label{fig:sigma} Left: 2D histogram of line strength and calibration precision is shown for Th lines at an operating current of 12\,mA and an exposure time of 10\,s. The coordinates, line strength (given in digital counts, i.e. analog-to-digital units) and RV precision [ms$^{-1}$], are given in log scale. The size of the (log) histogram bins is given by the little square right next to the grey-scale bar. The grey-scale bar indicates the number of lines per bin (linear scaling). The saturation limit is indicated by the horizontal dashed line. Right: The usage of many lines enhances the RV precision. Here, lines of similar precision (in bins of 0.1\,dex) are combined to enhance the RV precision. For each group of lines in a bin, the combined RV precision (left axis) is displayed against the precision for an individual line of that bin (bottom axis). The line of identity is shown in addition.}
\end{figure*}

In the following, we consider the calibration of measurements of radial velocities (RV). Since stellar line shifts are measured w.r.t. the calibration spectrum, the position of calibration lines has to be determined very precisely so that overall RV precision is not limited by the calibration spectrum. The impact of line shape and strength on the achievable RV precision can be predicted using the formulae given by \citet{Butler+1996} and \citet{Bouchy+2001}. The result is that there are many faint lines with low precision and a smaller number of strong lines with high precision (see the panel to the left in Figure~\ref{fig:sigma} for an example). The precision improves with line intensity and becomes worse the higher the noise is. The left-hand panel in Fig.~\ref{fig:sigma} shows that we need lines with thousands of counts to get a precision of tens of \mps. But the choice is not necessarily limited to the strongest lines. Although the position of a fainter line can be measured with lower accuracy only, the combination of a large number of them can statistically improve the precision of the wavelength calibration.

Therefore, in practice, all lines are combined to increase the RV precision. Here, we combine groups of lines having the same level of precision (collect lines in the same bin of precision and divide this value of precision by the square root of the number of lines in that bin; panel to the right in Fig.~\ref{fig:sigma}). Then, a combination of many strong lines allows one to get to a precision of 1\,\mps or better. For example, taken both panels of Fig.~\ref{fig:sigma} together, the combination of strong lines at a precision of 10\,\mps (in a small bin of only 0.1\,dex) for a current of 12\,mA and an exposure time of 10\,s results in a combined RV precision of a few \mps.

It is not possible in this work to show the actual final precision achievable with ThNe lines since this depends clearly on the particular setup and methodologies of RV surveys. However, our study highlights another aspect which is inherent to the ThNe spectrum. While the highest combined precision is achieved with strong lines at high operating current or long exposure time, the right-hand panel of Fig.~\ref{fig:sigma} also shows that the opposite is true at the faint end. There, the highest precision is achieved for short exposures and low current, although it is still much worse than the precision achieved with strong lines. The reason becomes clear from an inspection of the distribution of line strength. In the next section we will see that the number of faint lines increases with decreasing operating current and exposure time, increasing the combined precision of faint lines.

\subsection{Number of Th lines and distribution of line strength}
\label{sect:strengths}

\begin{figure}
\includegraphics[width=\hsize]{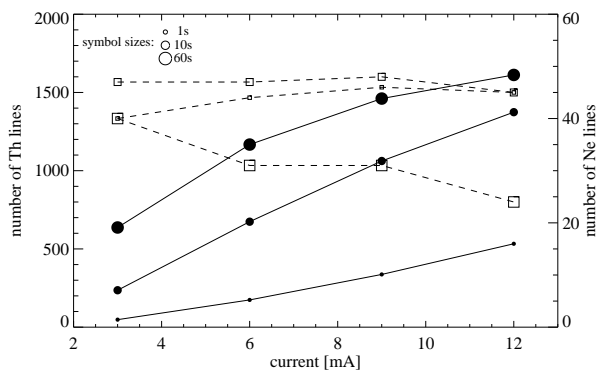}
\caption{\label{fig:plot_numbers2} The number of Th (circles) and Ne lines (squares) vs. operating current. Only lines above the noise level (100\,counts) and below the saturation limit (5.4 in log. scale) were included. Symbol size indicates the value of exposure time as is shown in the legend. Note that there are different axes for Th and Ne lines.}
\end{figure}

The previous assessments make clear that the number of strong Th lines needs to be maximised. At the same time, the number of saturated lines should be kept at a low level. Therefore, for each spectrum, Th and Ne lines were counted (Fig.~\ref{fig:plot_numbers2}).

The number of Th lines increases significantly with increasing exposure time and operating current. The increase with operating current is close to linear but flattens at long exposure time. The number increases non-linearly with exposure time, notably with opposite behaviour at low and high currents. The number of Ne lines behaves differently. It is close to constant for short and intermediate exposures but decreases with increasing current for long exposures.

\begin{figure}
\includegraphics[width=\hsize]{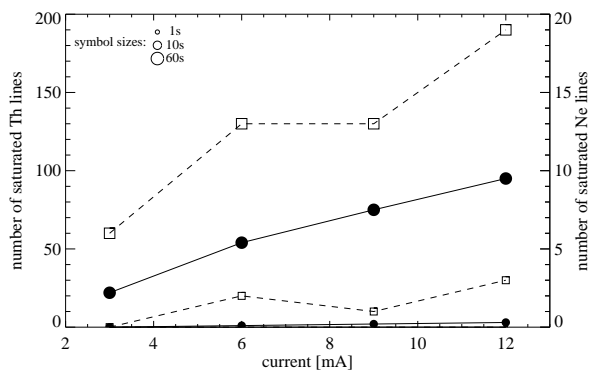}
\caption{\label{fig:plot_sat2} The number of saturated Th (circles) and Ne lines (squares). The layout follows Fig.~\ref{fig:plot_numbers2}.}
\end{figure}

The number of saturated lines rises quickly with increasing exposure time (Fig.~\ref{fig:plot_sat2}). There is also a strong increase with operating current for intermediate and long exposure times. Saturation is essentially absent in short exposures\footnote{The 63 strong Ne lines are excluded from this analysis.}. The increase of saturating Ne lines reflects the correct magnitude of the number of Ne lines disappearing in Fig.~\ref{fig:plot_numbers2} when operating current and exposure time increase. Those are the lines entering the regime of saturation at  higher current and longer exposure time. The flattening of the Th relation cannot be explained by saturation as the increase of saturating Th lines (max. $\approx100$) is not high enough.

\def \imgwidth{5cm}
\begin{figure*}
\center
\begin{tabular}{ccc}
\subfigure{\includegraphics[width=\imgwidth]{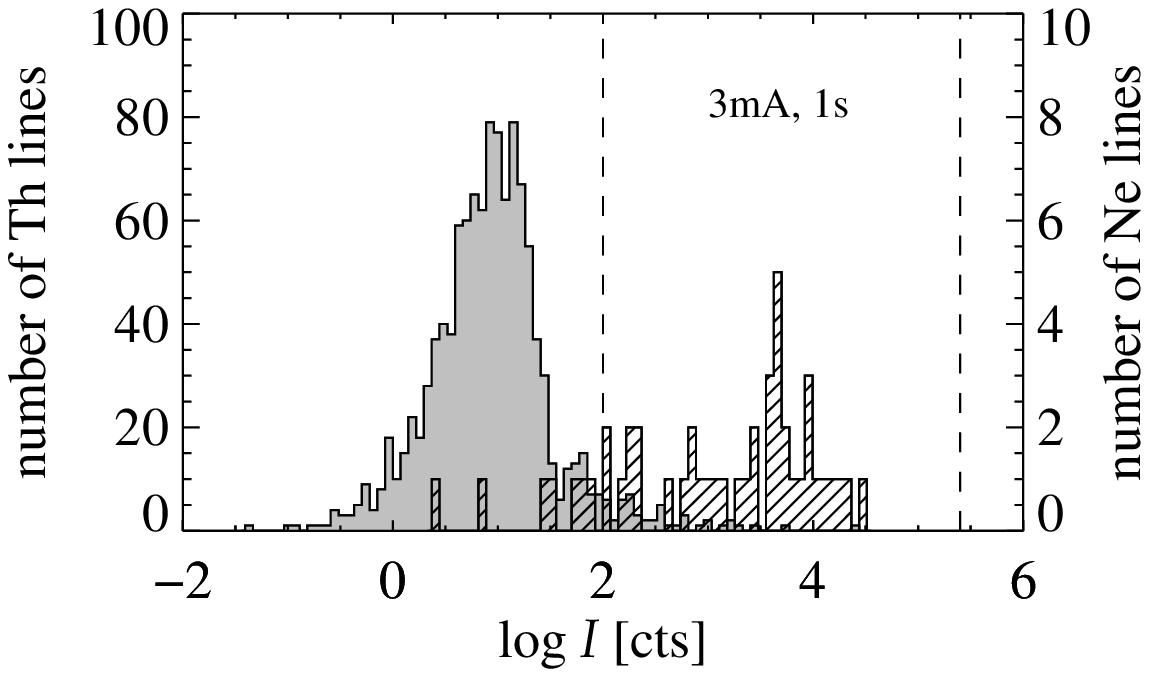}}&
\subfigure{\includegraphics[width=\imgwidth]{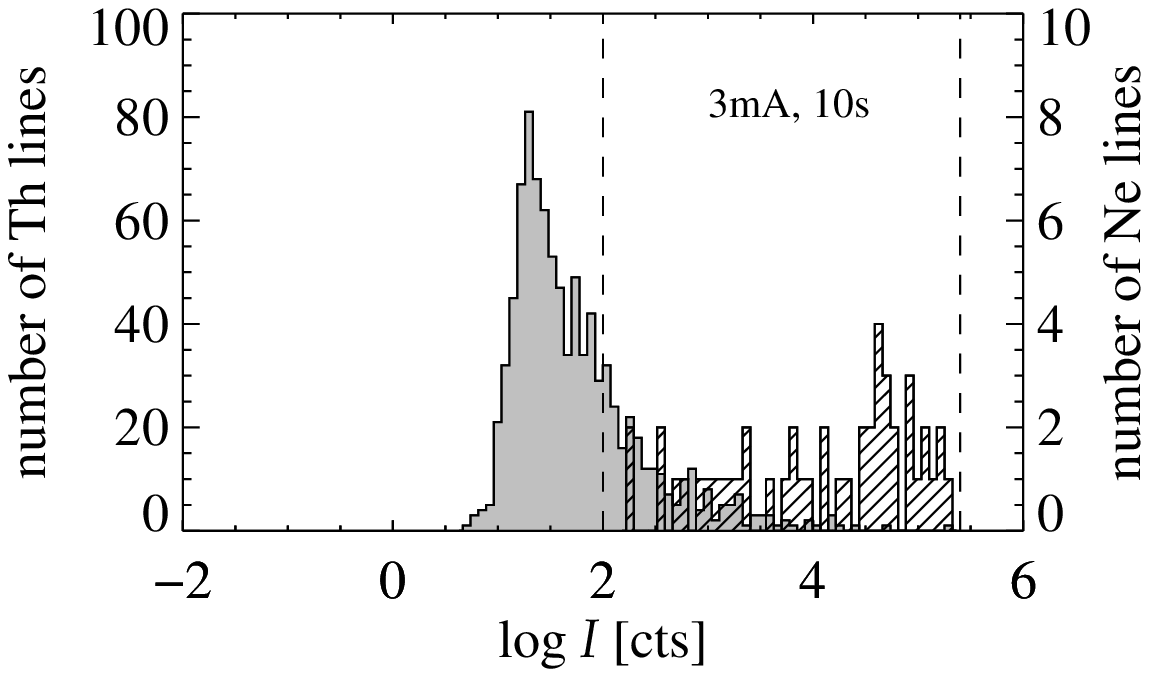}}&
\subfigure{\includegraphics[width=\imgwidth]{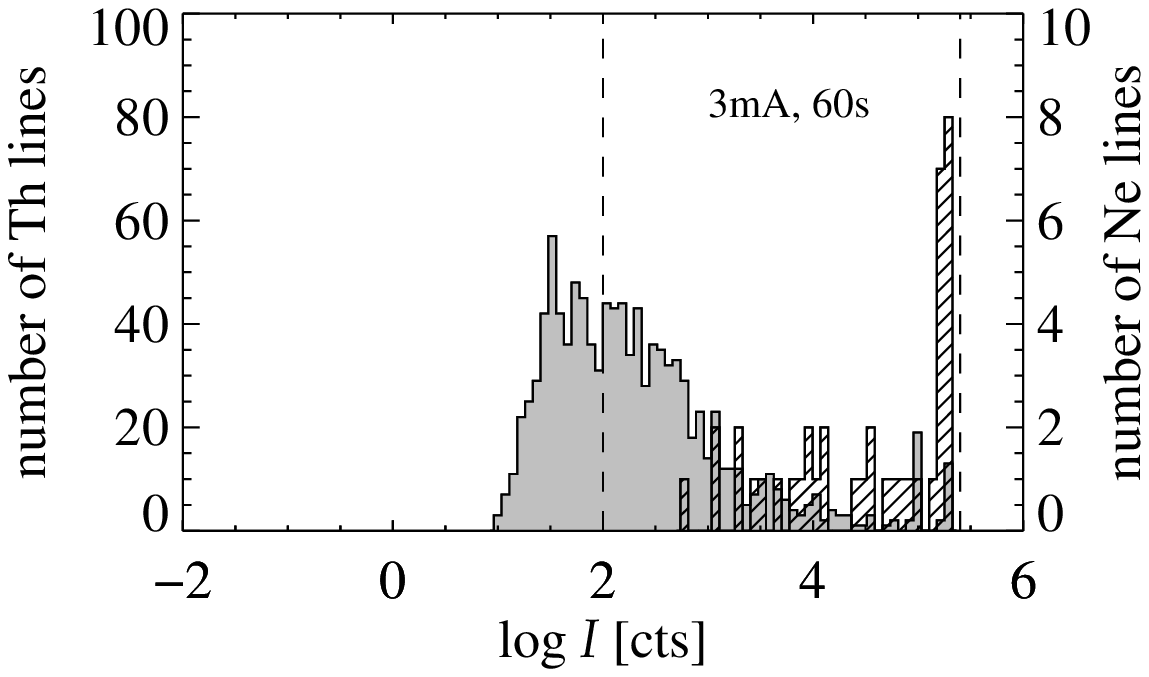}}\\
\subfigure{\includegraphics[width=\imgwidth]{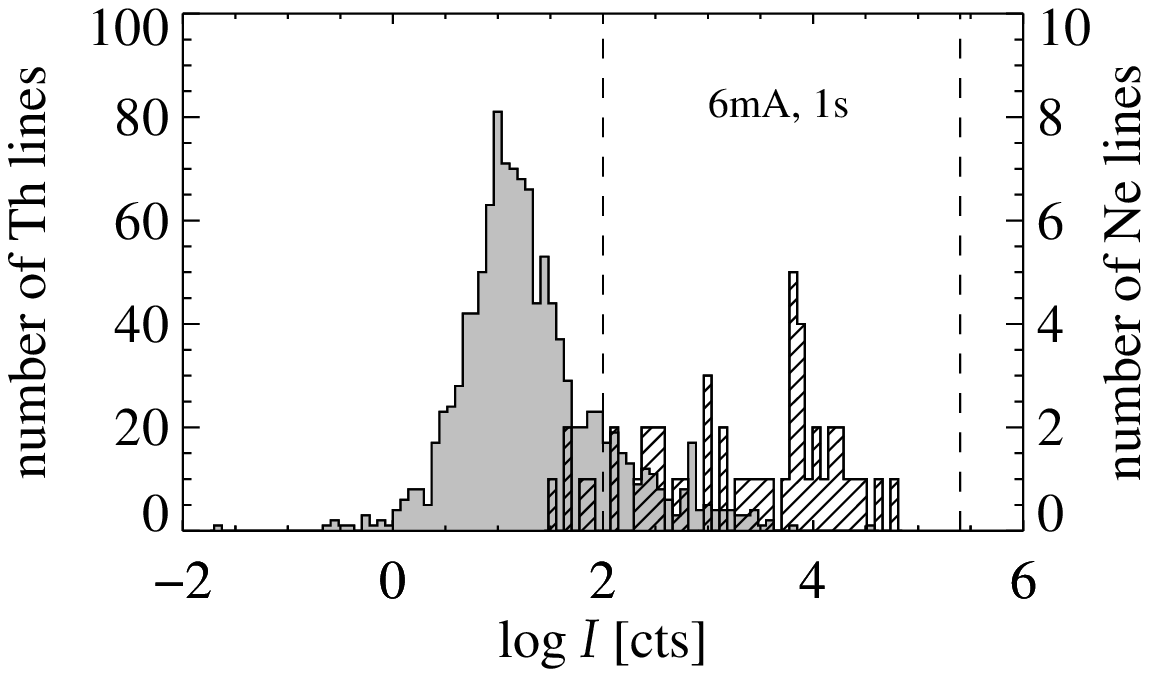}}&
\subfigure{\includegraphics[width=\imgwidth]{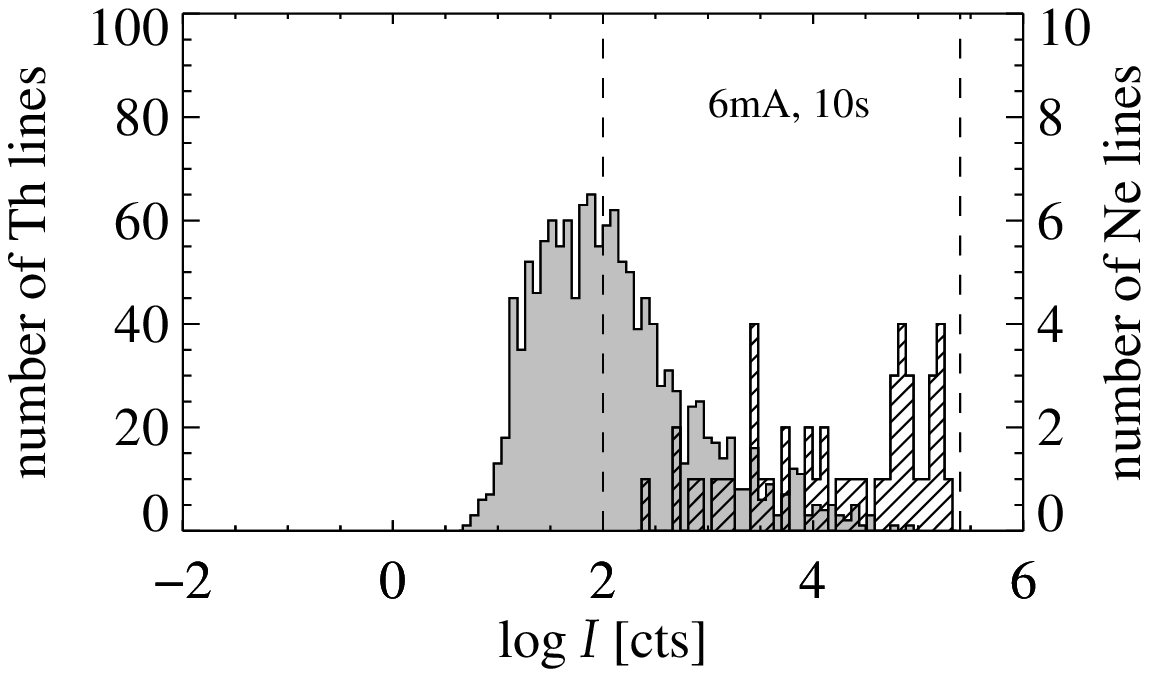}}&
\subfigure{\includegraphics[width=\imgwidth]{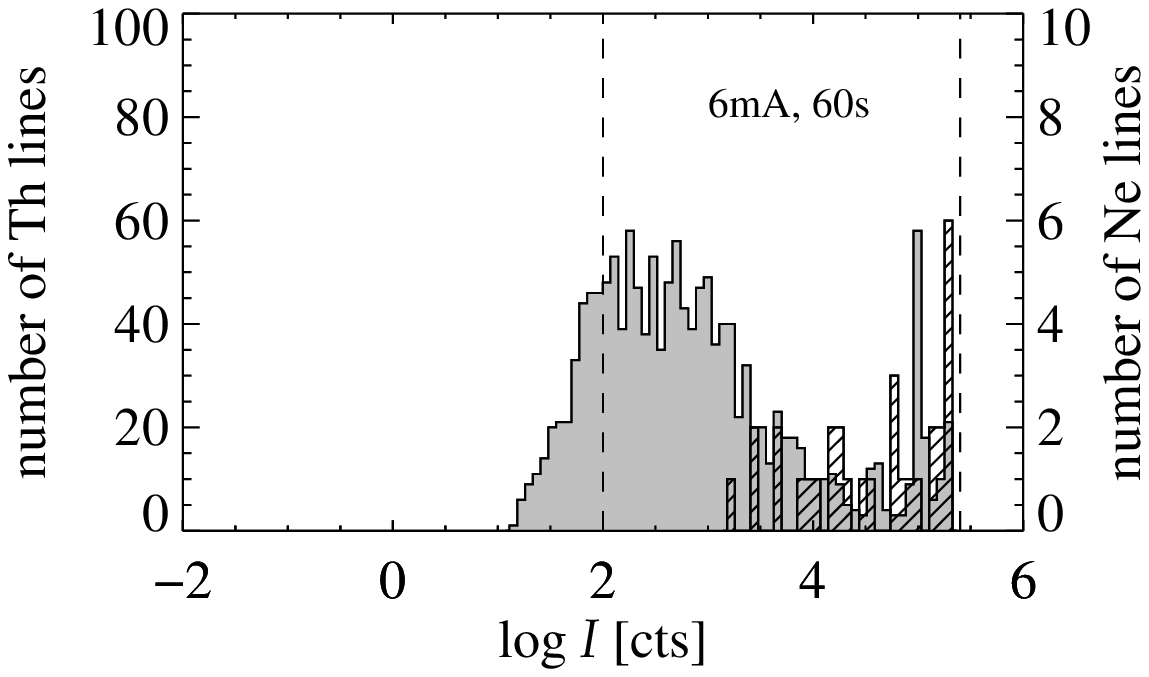}}\\
\subfigure{\includegraphics[width=\imgwidth]{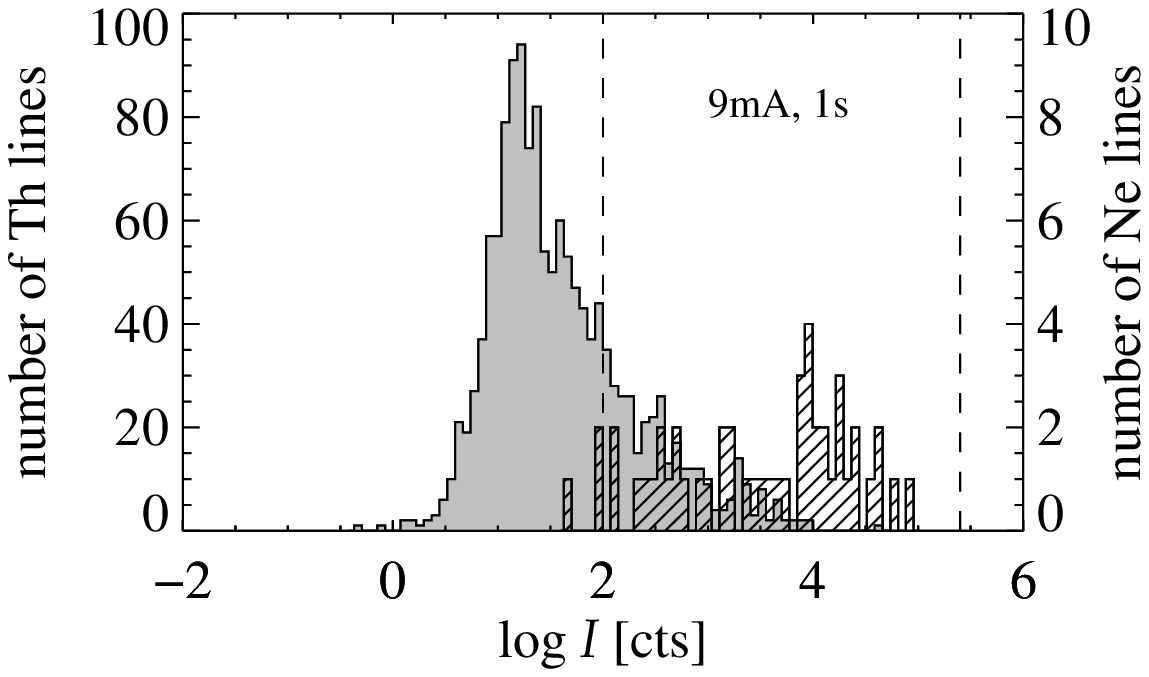}}&
\subfigure{\includegraphics[width=\imgwidth]{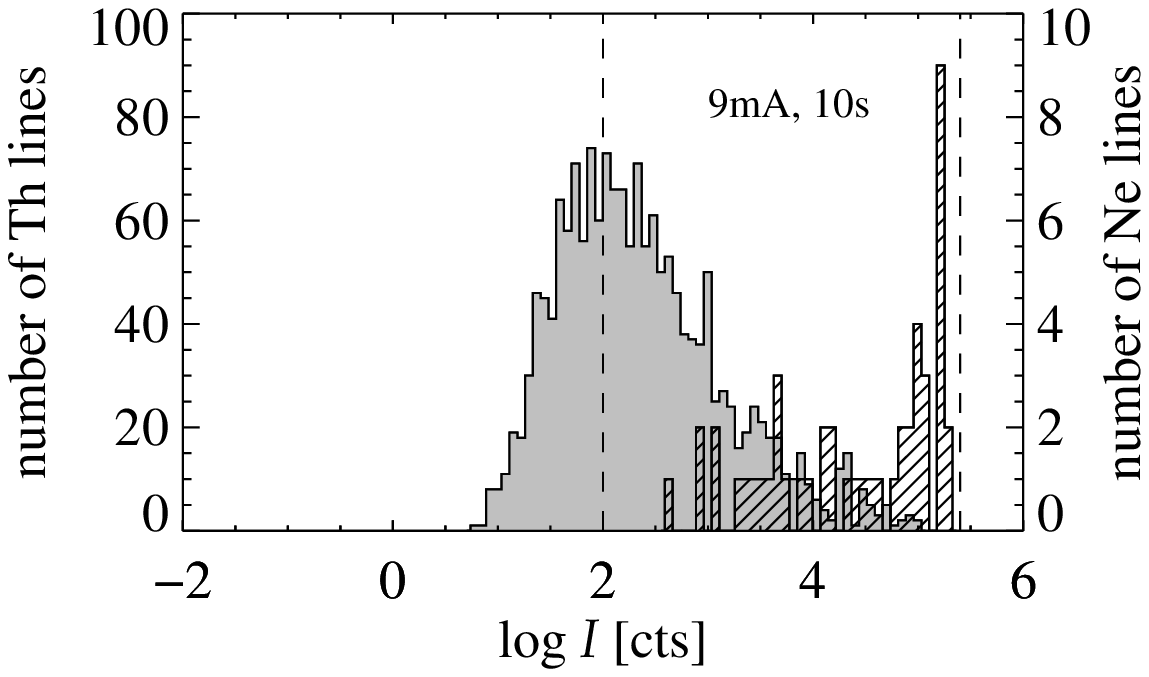}}&
\subfigure{\includegraphics[width=\imgwidth]{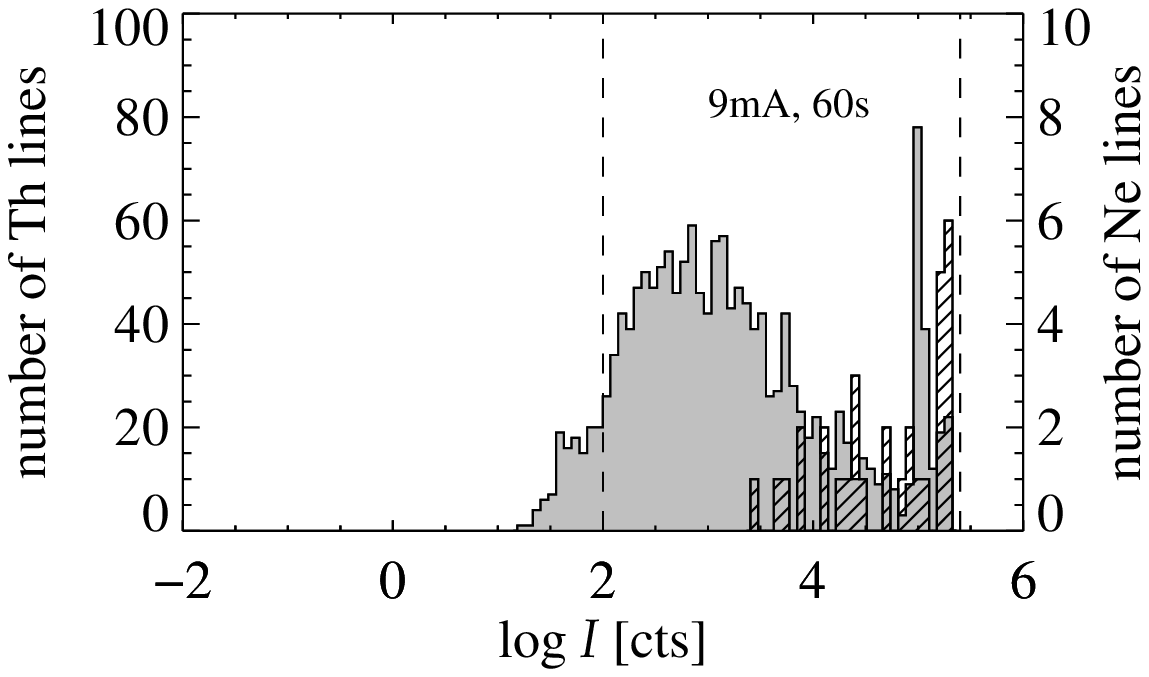}}\\
\subfigure{\includegraphics[width=\imgwidth]{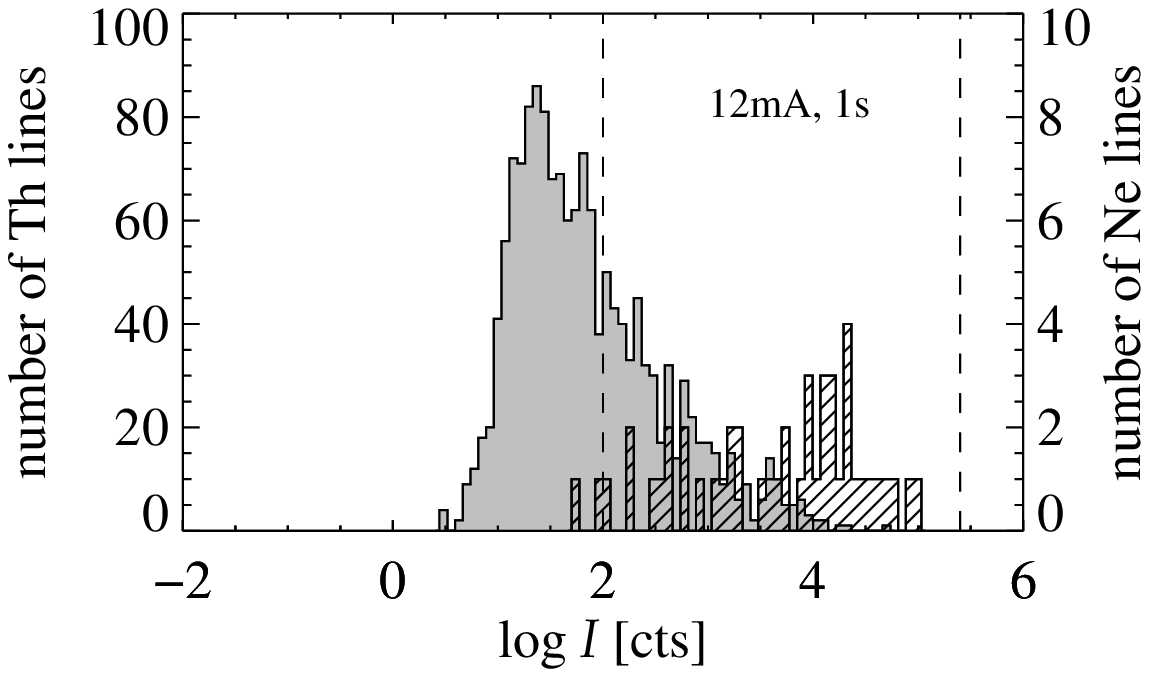}}&
\subfigure{\includegraphics[width=\imgwidth]{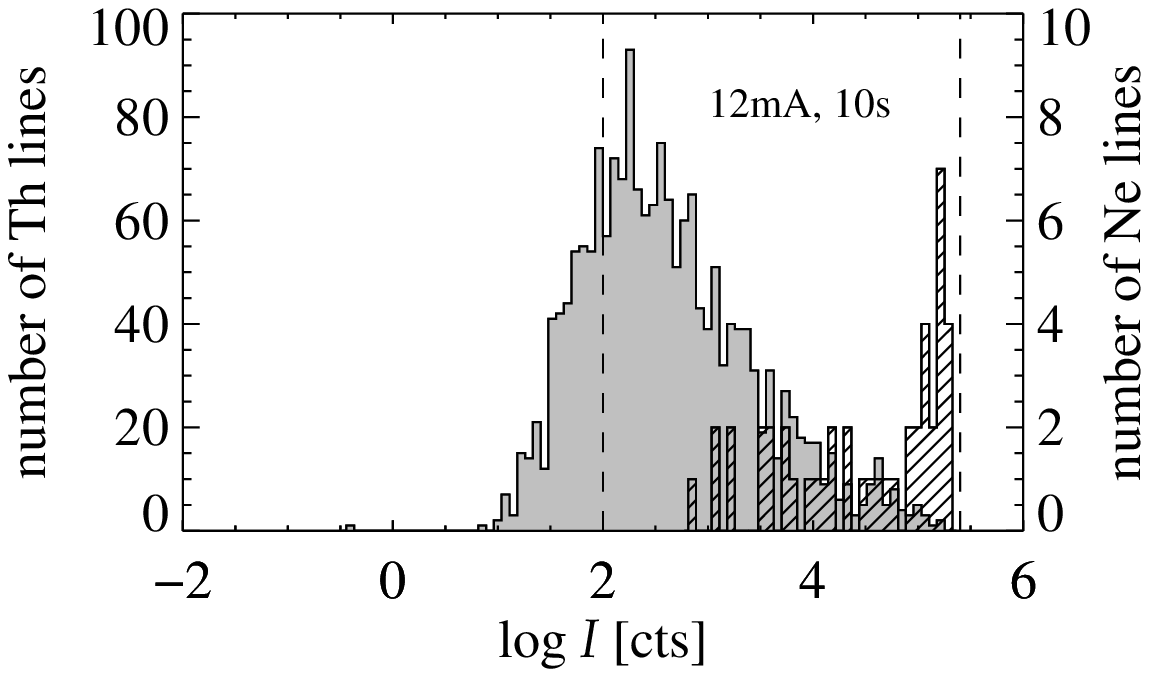}}&
\subfigure{\includegraphics[width=\imgwidth]{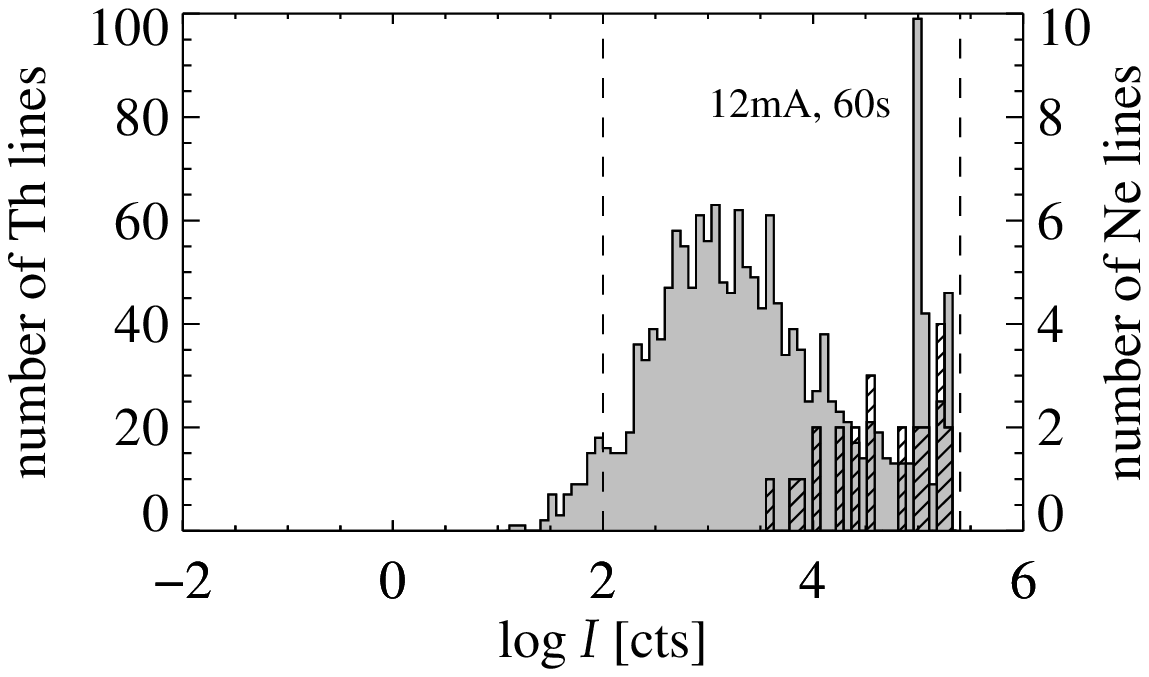}}
\end{tabular}
\caption{\label{fig:stat_strength2} The histograms show the distributions of line strengths of Th lines (grey) and Ne lines (hatched) taken at different combinations of operating current (rows) and exposure times (columns). Dashed lines indicate the noise and saturation limits adopted for various analyses in the present work. A mask has been created from the 12mA, 10s spectrum to exclude saturated columns from the analysis. This mask could by far not cover the saturated columns in the spectra with long exposures of 60\,s, indicated by the high peaks at a log. line strength of 5.0 and 5.2.}
\end{figure*}

The distribution of line strength for each spectrum sheds some more light on the behaviour of line number (Fig.~\ref{fig:stat_strength2}). First of all, we notice the single-peaked distribution of Th lines with similar height and width (in log. space) for all settings. While the distributions shift swiftly with exposure time, there is also a slight shift with operating current. An increase of operating current from 3 to 12\,mA shifts the peak of the distribution by one order of magnitude in a rather linear way. This is reflected by the well-known brightening of the lamp when operating current is increased.

This explains the line numbers observed. At low current, an increase of exposure time lifts first the tail and then the bulk of the distribution above the noise level. Therefore, line number increases non-linearly with positive curvature. At high current, the increase of exposure time lifts the bulk and then the tail above the noise level. There, the relation flattens with increasing exposure time. At moderate current, an increase of exposure time samples the central part of the distribution in a rather symmetric way. Therefore, the increase of line number with exposure time is close to linear.

For Ne lines we find that the distribution is rather uniform with a small peak with intensities about 3 orders of magnitude higher than the Th peak intensity. The distribution of Ne lines hardly changes when operating current is increased. Therefore it does not come as a surprise that the number of unsaturated Ne lines hardly changes with operating current and is strongly affected by lines entering the saturation regime.

The distribution of line strengths favours a choice of 9 or 12 mA with an exposure time of 10s to optimise the number of suitable Th lines. At long exposure times and high values of operating current, most Th lines are above the approximate noise level (about 100\,cts) and below the saturation limit. At an exposure time of 10\,s, the spectra taken with 9\,mA and 12\,mA display a high number of useful Th lines and a low level of saturation at the same time. At these operating conditions, the benefits for Th are not impaired by the Ne lines which hardly vary at moderate and long exposure times. The choice of 9 mA instead of 12mA and a choice of 10s instead of 60s further optimise the life time of the lamp.

\section{Summary and Conclusions}
\label{sect:concl}

Spectra of a ThNe lamp were obtained from $5,440\,${\AA} to $10,250\,${\AA} with a setup in Tautenburg using different exposure times and operating currents. The particular values of exposure time were optimised for the setup in Tautenburg and will have to be adjusted for a repetition of the measurements at other spectrographs. Of course, the exact choice of exposure times has no effect at all on the general conclusions of the present work. 

A list of $\ga$2,000 Th lines was identified using known wavelengths from published line lists. We did not attempt to find all available lines. Rather, the goal was to identify lines in a homogeneous way for all spectra taken with different settings.

An atlas of a ThNe spectrum has been produced identifying 63 Ne lines which show strong damping wings. They affect more than 300 Th lines in their neighbourhood so that those are excluded from most analyses in the present study.

A comparison at different exposure times shows that lines with a peak intensity of less than 100 digital counts above the background will be affected by noise. The saturation limit is identified at 5.4 in log. units of digital counts.
The number of suitable Th lines, i.e. unsaturated isolated lines above the noise limit, increases strongly with exposure time and operating current. The increase can be explained by the distribution of line strengths and how this distribution is shifted towards higher intensities with increasing current and exposure time. The number of useful Th lines is hardly affected by saturation in the parameter regime studied.

\citet{2007ASPC..364..461K} emphasised that the intensities of the lines of the filling gas behave in a way completely different from Th so that they can be identified easily and separated from the Th lines. Indeed, Figs.~\ref{fig:12_6} and \ref{fig:plot_numbers2} show that Ne and Th lines change with operating current in distinct ways on average. Considering the spread of Th line strengths though (Fig.~\ref{fig:12_6}), there is some doubt whether individual lines of the filling gas can be identified this way. We expect that the change of the mean intensity ratio of the Th compared to the Ne lines is a general property of these lamps. We thus expect that a similar behaviour will be found at other wavelengths.

{Despite thousands of lines identified in each setting, there are only some 200 lines common to all spectra. Line confusion is unlikely to occur at such an extent. Instead, our study rather shows that the dynamic range is very large when changing exposure time and operating current. In other words, we assume that another set of lines appears at another value of operating current.
 
Although there is no systematic average trend with changing operating current, we cannot exclude variations in the wavelengths of Th lines due to unknown blends at a level that corresponds to radial velocity variations of hundreds of \mps. This finding indicates that a stable operating current of the calibration lamp can be necessary. In the present work, the measurement of possible correlations of wavelength residuals with operating current is hampered by the low number of data points (3 in exposure time and 4 in current). With spectra taken at additional values of operating current, the Spearman rank correlation coefficient would reveal systematic shifts of individual lines. Such trends with operating current can reveal line blends because of the relative change of the strength of the spectral line components.

Since wavelength variations found at this order of magnitude are at variance with previous work, a repetition of the analysis at very high spectral resolution with, e.g., a Fourier Transform Spectrometer is highly recommended.  Subtle changes of the position of individual lines can only be unveiled with a stabilised spectrograph, or if spectra of much higher spectral resolution are available that show whether a line is a blend or not.

The warm-up phase of the lamp might be critical if stable operating conditions turn out crucial. Experience with our ThNe lamp shows that voltage stabilises at a level of a few percent after 20\,min only, depending on the choice of operating current. The behaviour of individual spectral lines during warm-up was not studied in this work.

We estimated the achievable precision of radial velocity measurements for individual Th lines and found that a combination of strong Th lines enables measurements at the level of 1\,\mps. The behaviour of the number of suitable Th lines and of the achievable calibration precision is defined by the distribution of line strength and how it varies with operating settings. We conclude that a high operating current is of advantage, at the expense of lamp life time and ageing effects. A high value of operating current reduces the residuals of wavelengths of identified lines and maximises the number of strong Th lines, which are necessary for a high radial velocity precision, without degradation by Ne lines. While strengthening the Th lines with a high operating current, the exposure time can be reduced to avoid saturation and blends, and to further reduce wavelength residuals.

\acknowledgements
This work was supported in part by DLR (Deutsches Zentrum f\"ur Luft- und Raumfahrt) under the grants 50OW0204 and 50OO1501. We would like to thank the workshops at the observatory in Tautenburg, Germany. In particular, we want to thank Johannes Winkler who built and adjusted the optomechanical setup in Tautenburg and Michael Pluto who was responsible for the electrical supply of the hollow cathode lamps. We thank R. Engleman for providing a digital version of the Th line list. We would like to thank the anonymous referee for the very helpful comments. This research has made use of NASA's Astrophysics Data System Bibliographic Services.

\newpage

\appendix

\section{ThNe atlas}
\label{app:A}
\begin{figure}
\includegraphics[width=\atlaswidth]{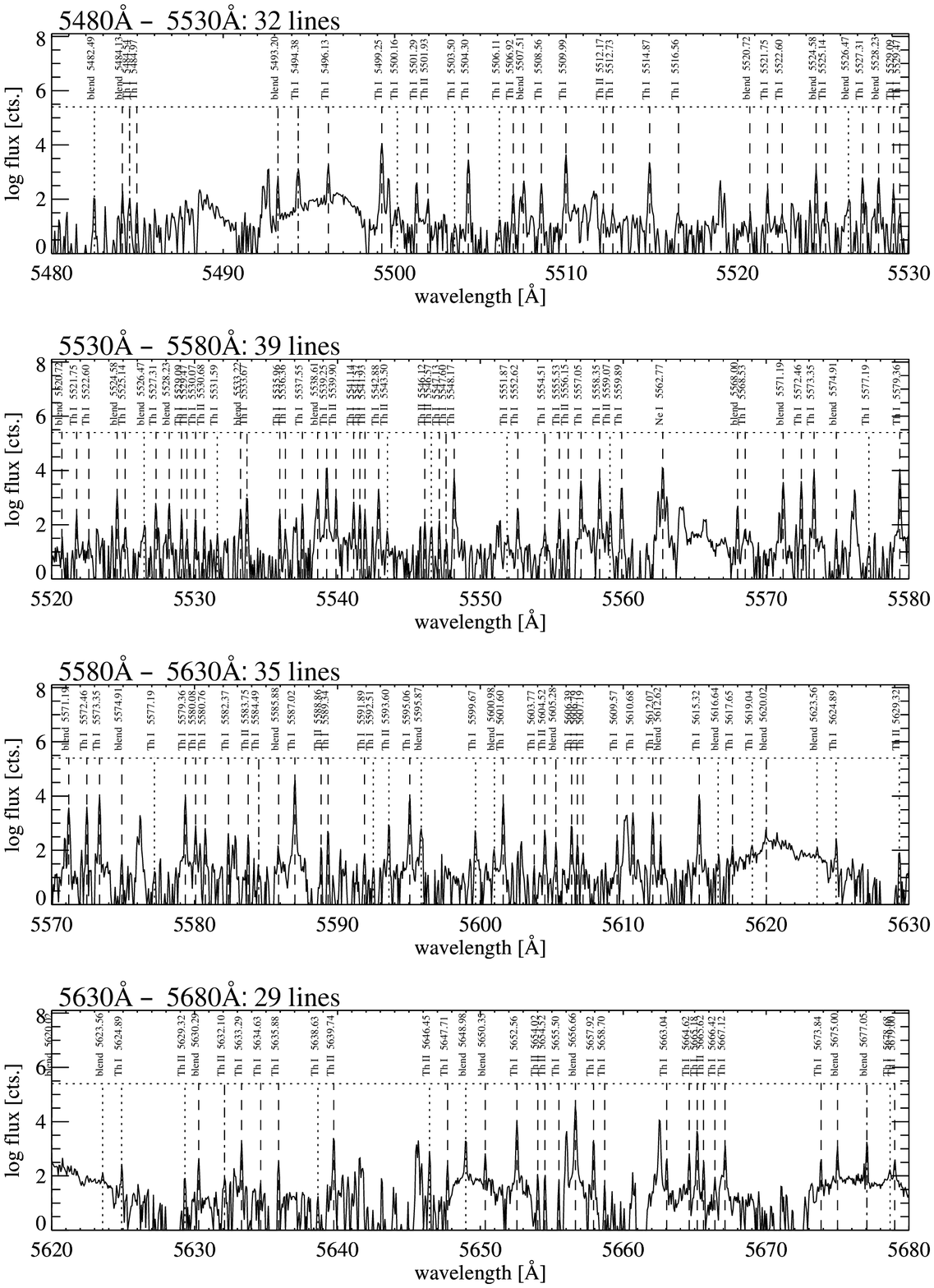}
\end{figure}
\clearpage
\begin{figure}
\includegraphics[width=\atlaswidth]{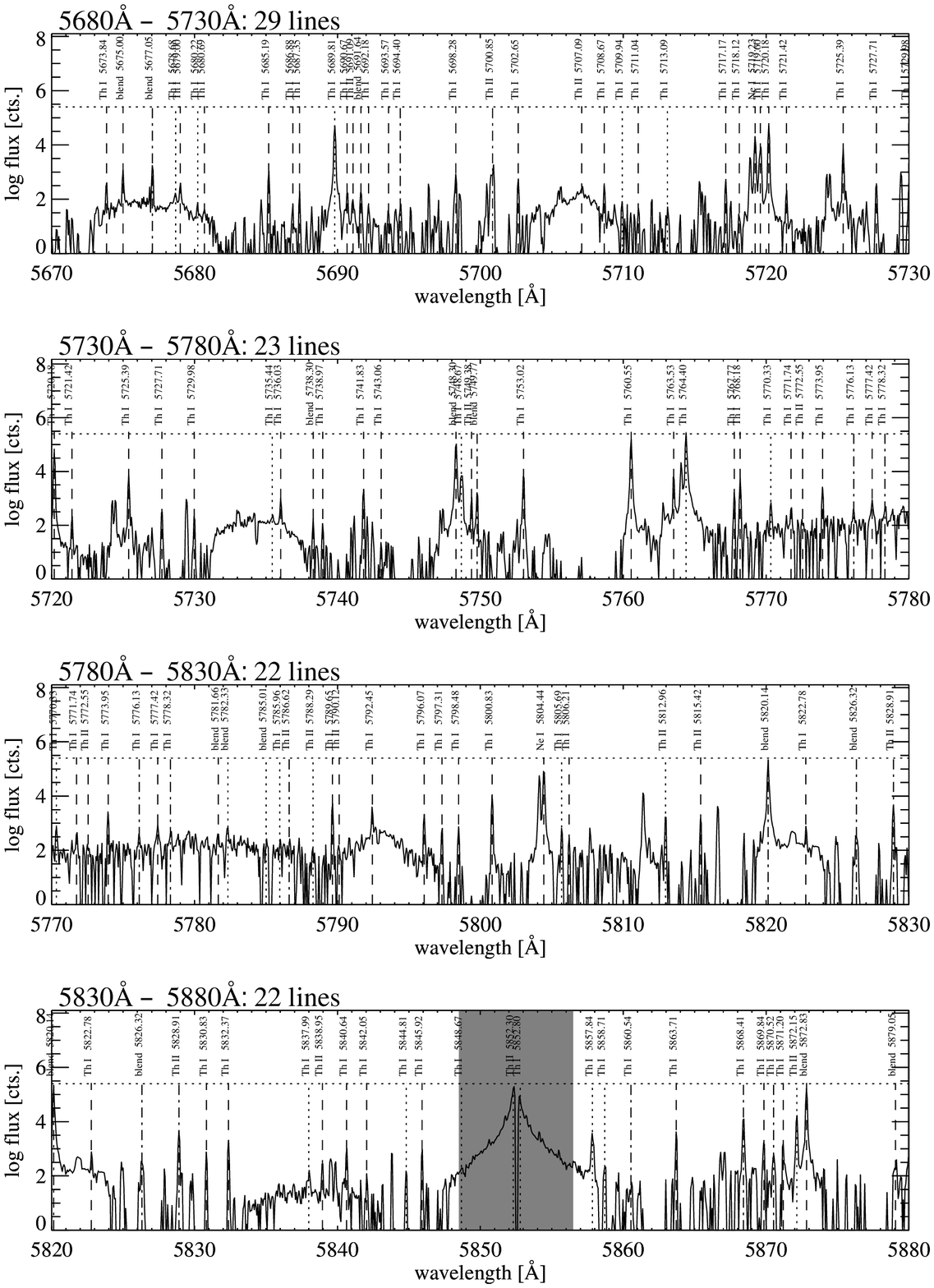}
\end{figure}
\clearpage
\begin{figure}
\includegraphics[width=\atlaswidth]{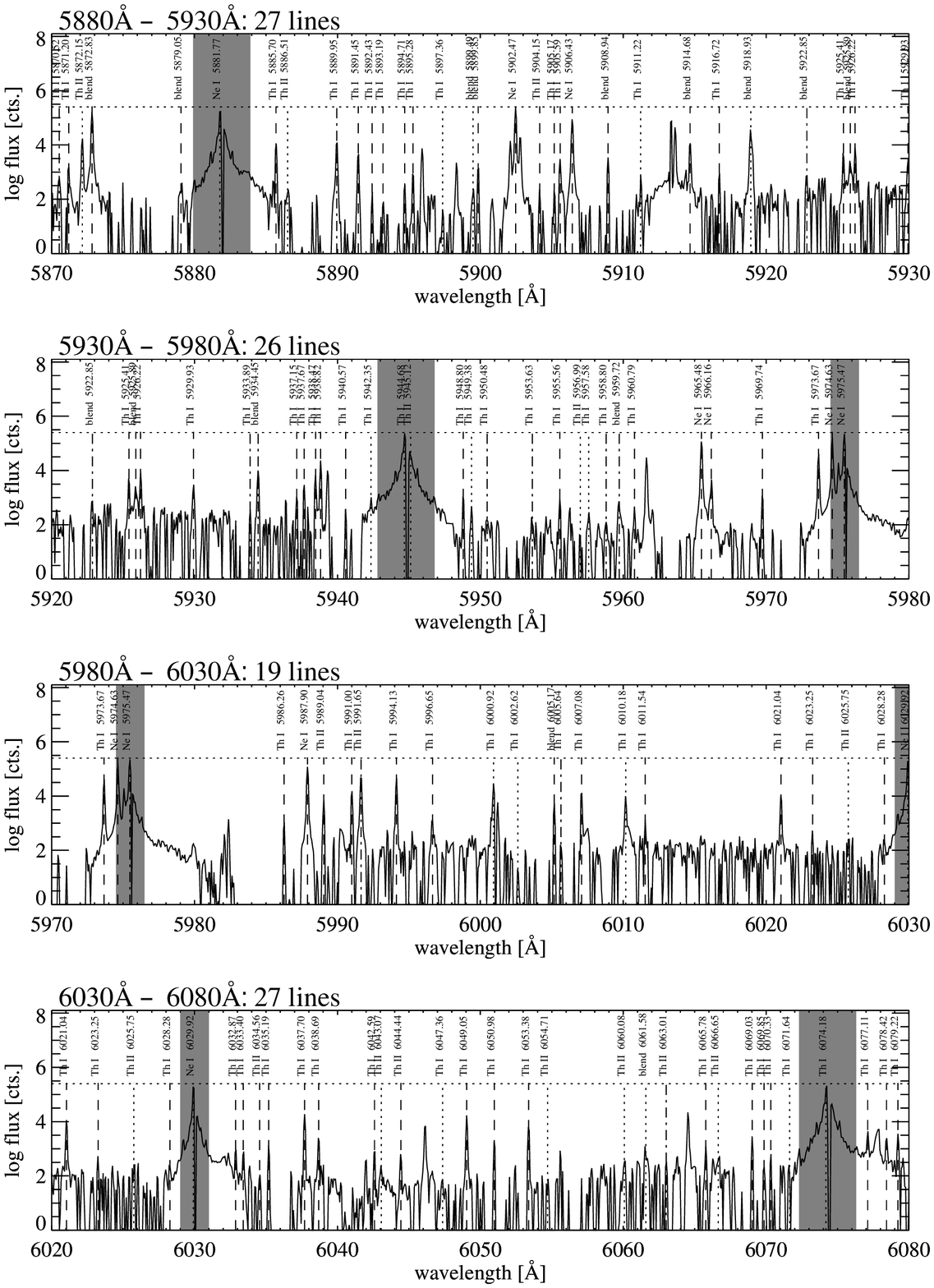}
\end{figure}
\clearpage
\begin{figure}
\includegraphics[width=\atlaswidth]{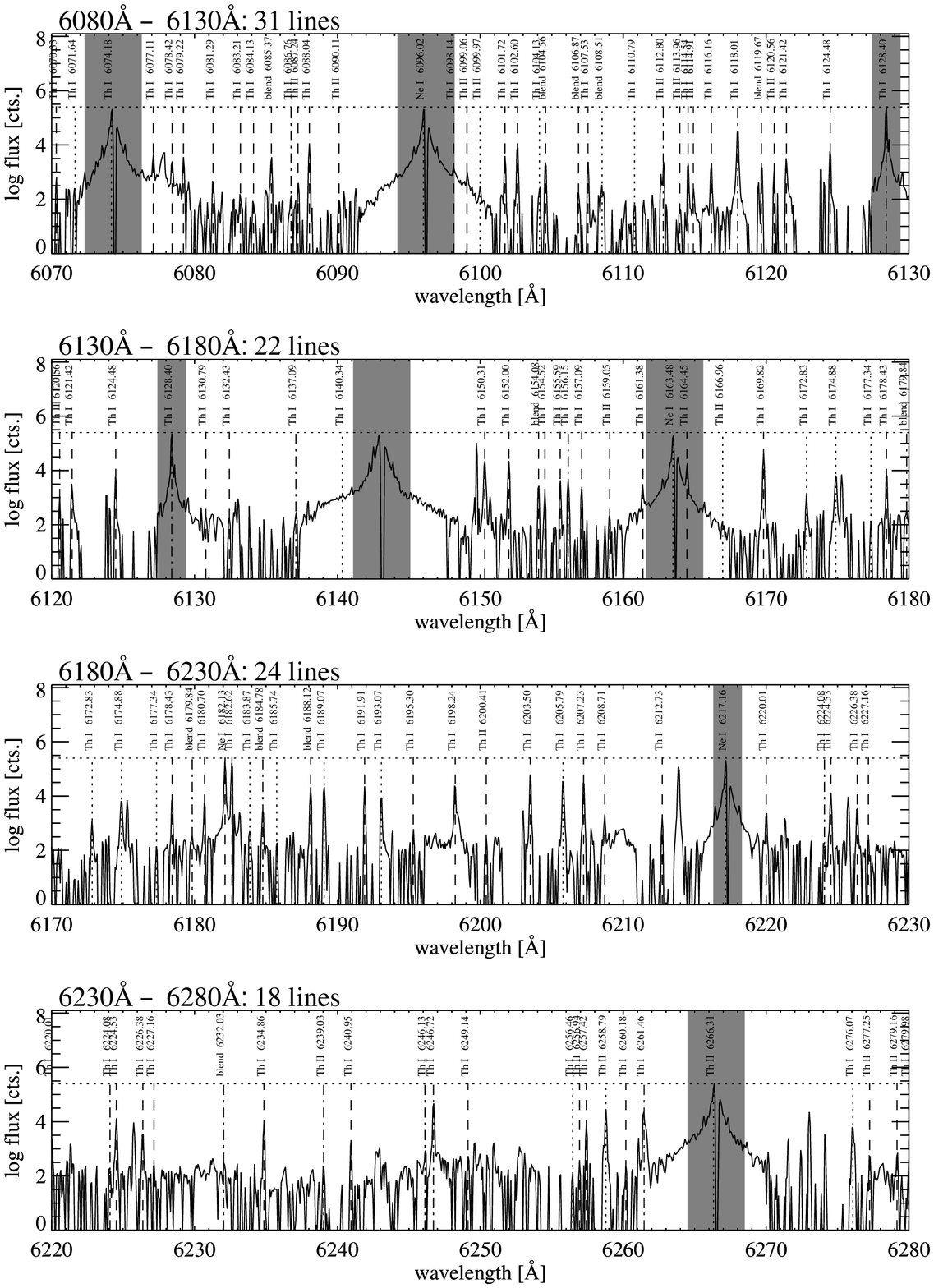}
\end{figure}
\clearpage
\begin{figure}
\includegraphics[width=\atlaswidth]{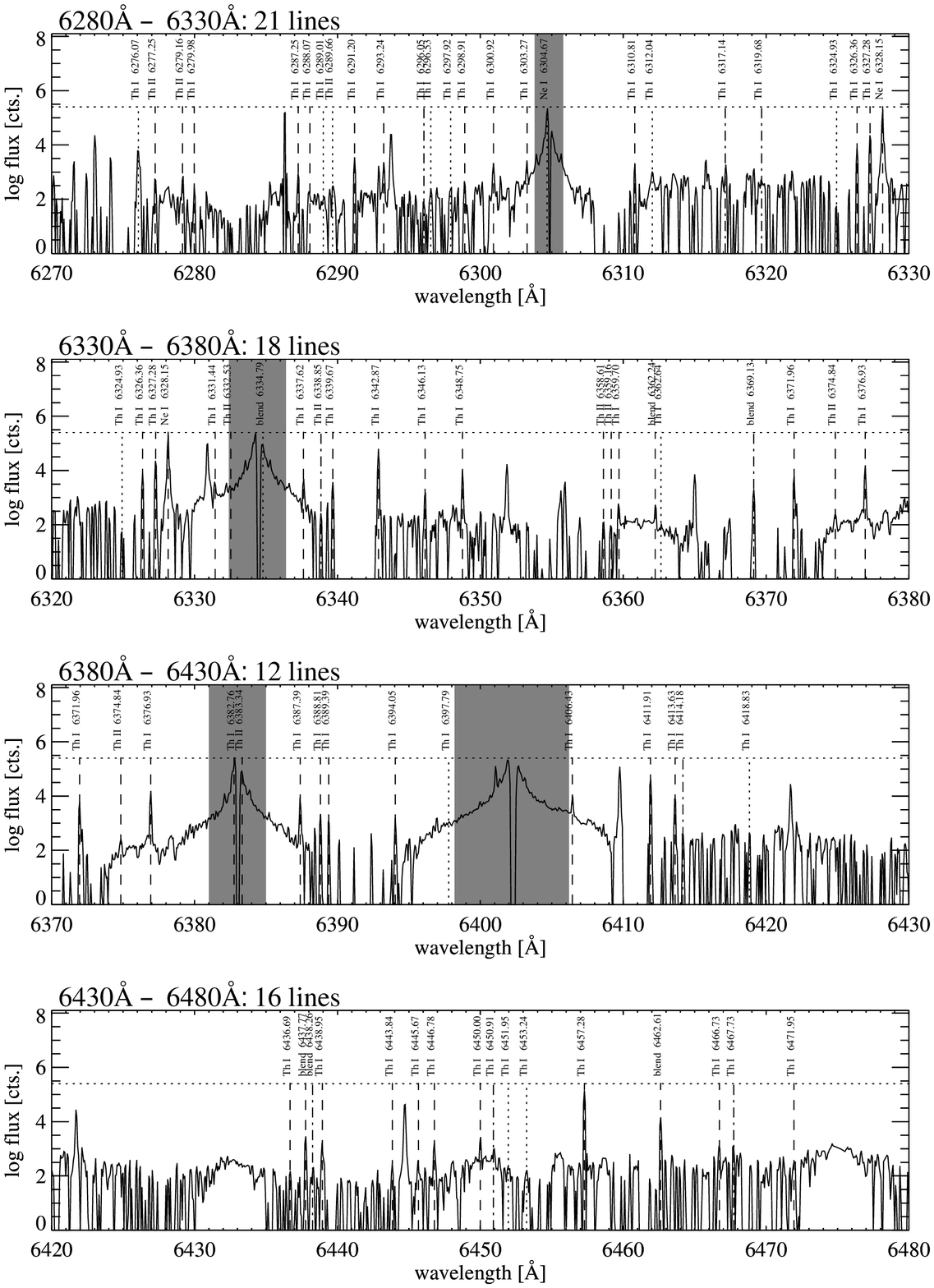}
\end{figure}
\clearpage
\begin{figure}
\includegraphics[width=\atlaswidth]{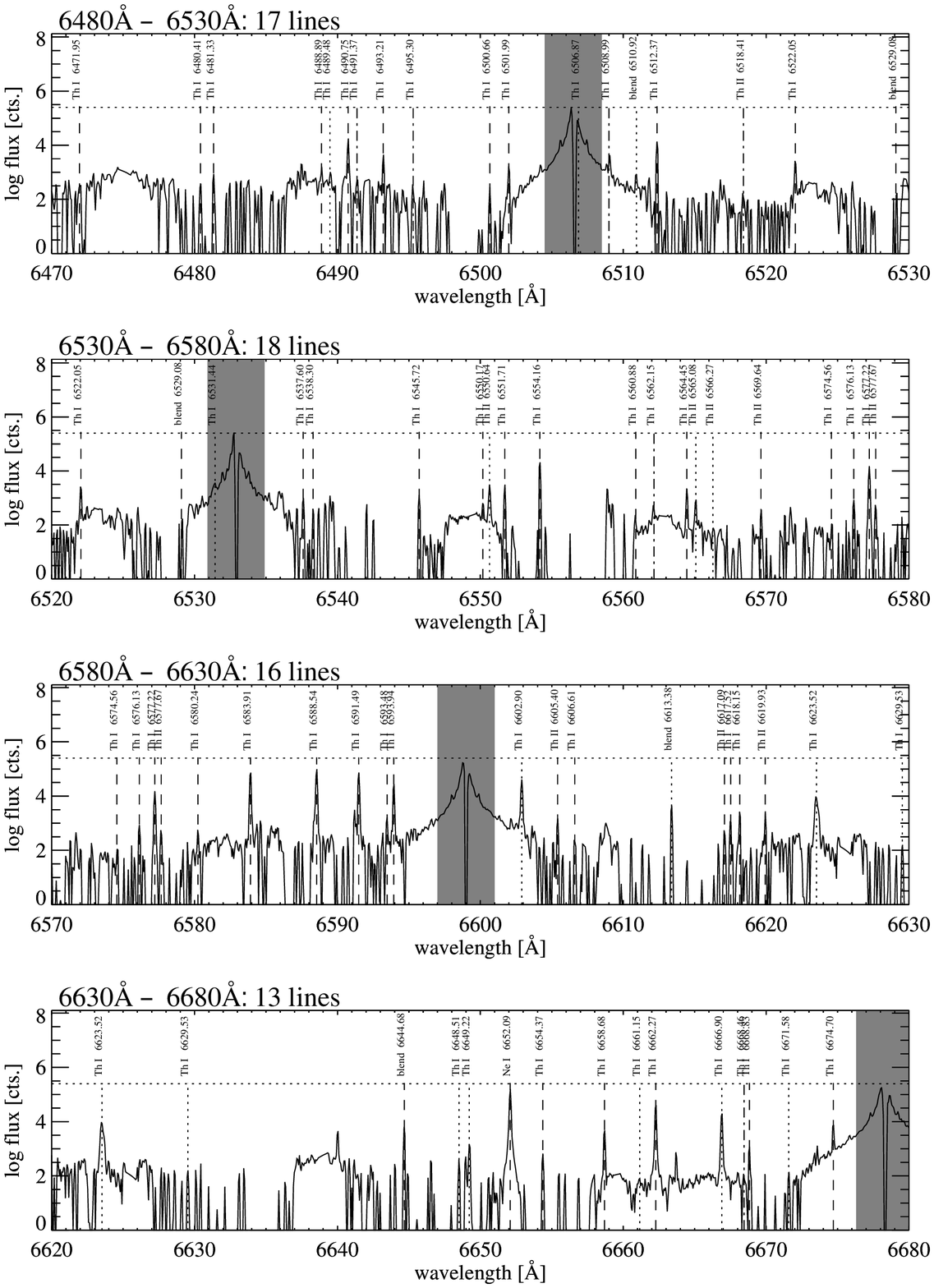}
\end{figure}
\clearpage
\begin{figure}
\includegraphics[width=\atlaswidth]{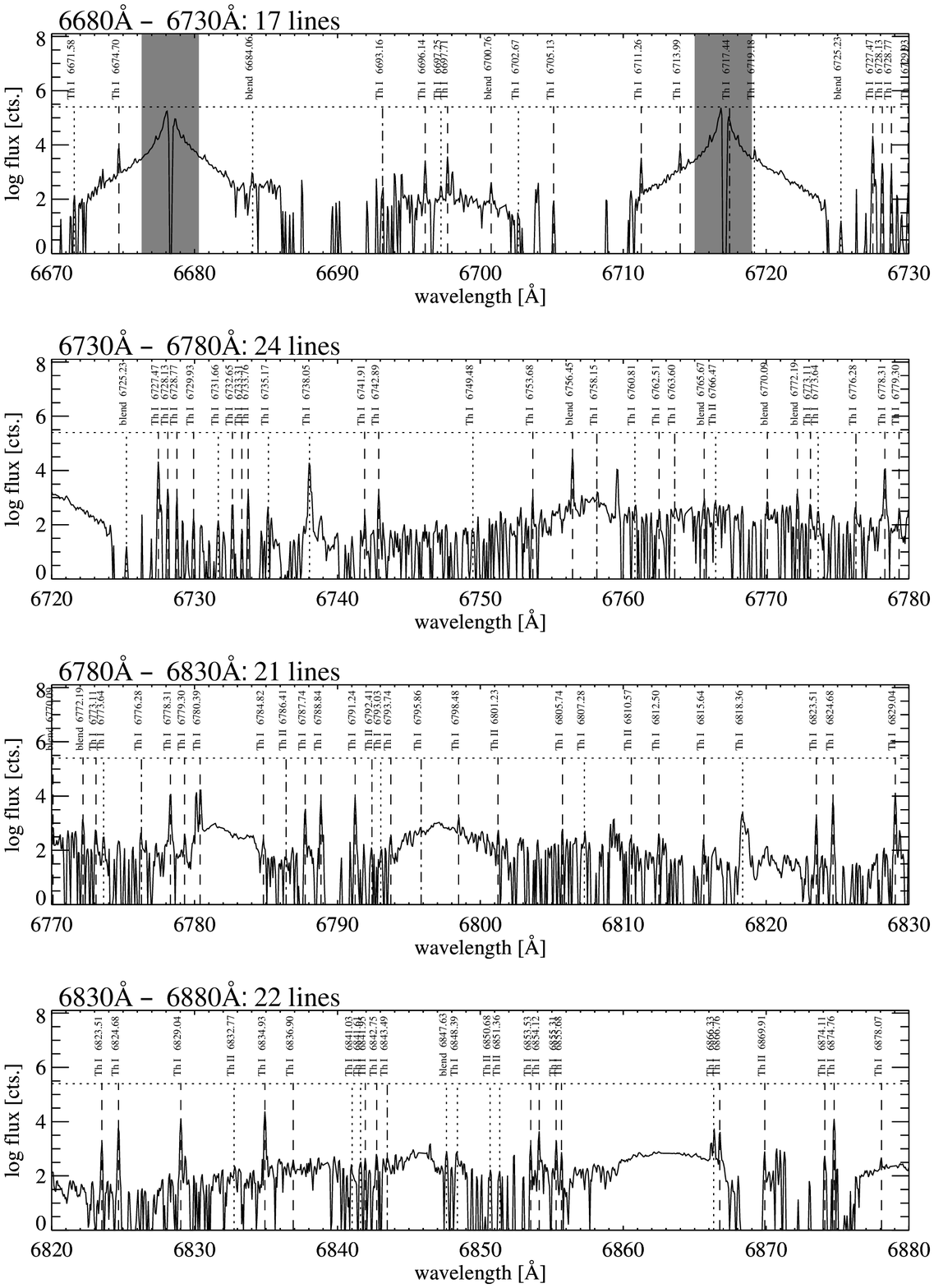}
\end{figure}
\clearpage
\begin{figure}
\includegraphics[width=\atlaswidth]{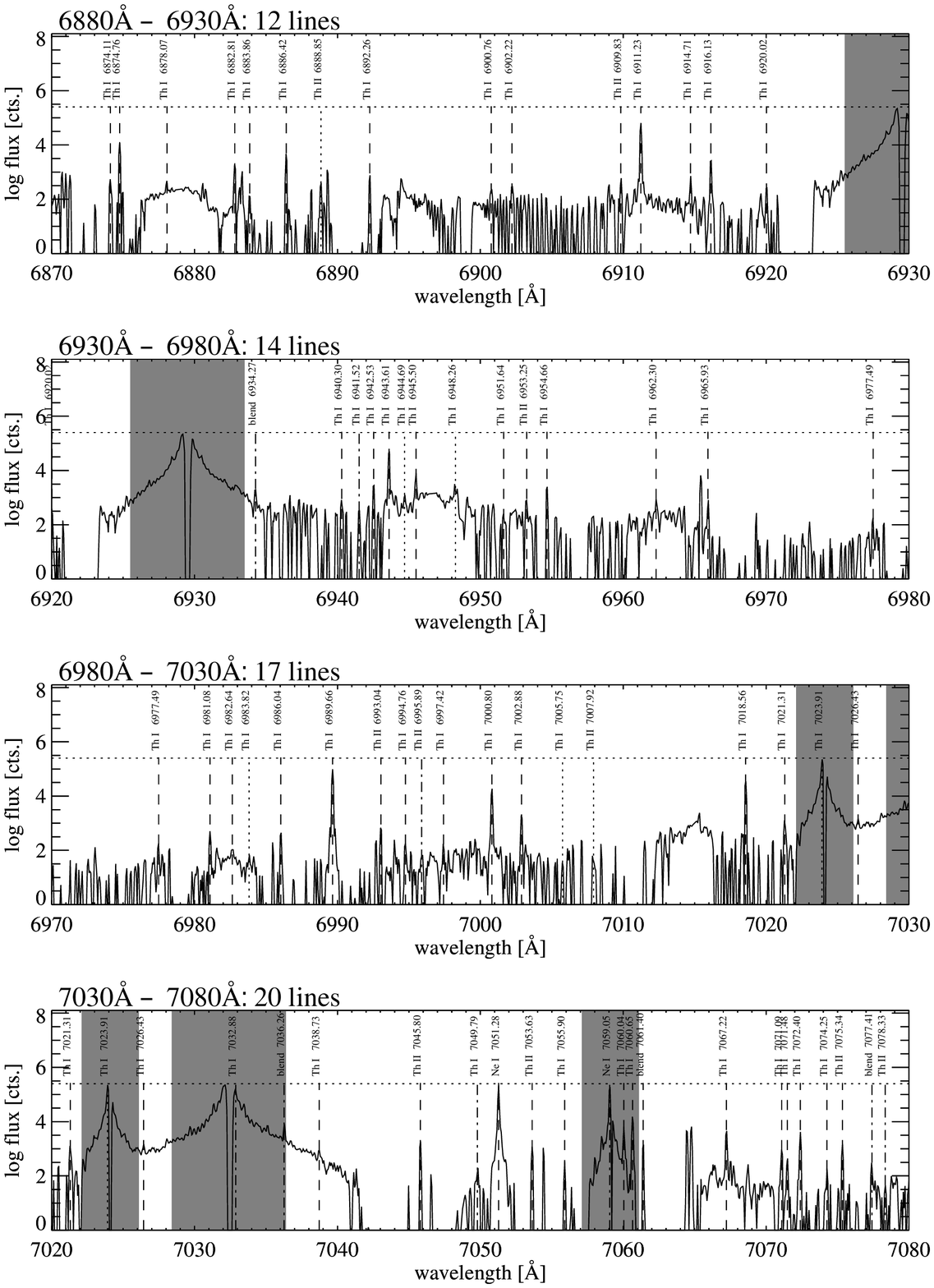}
\end{figure}
\clearpage
\begin{figure}
\includegraphics[width=\atlaswidth]{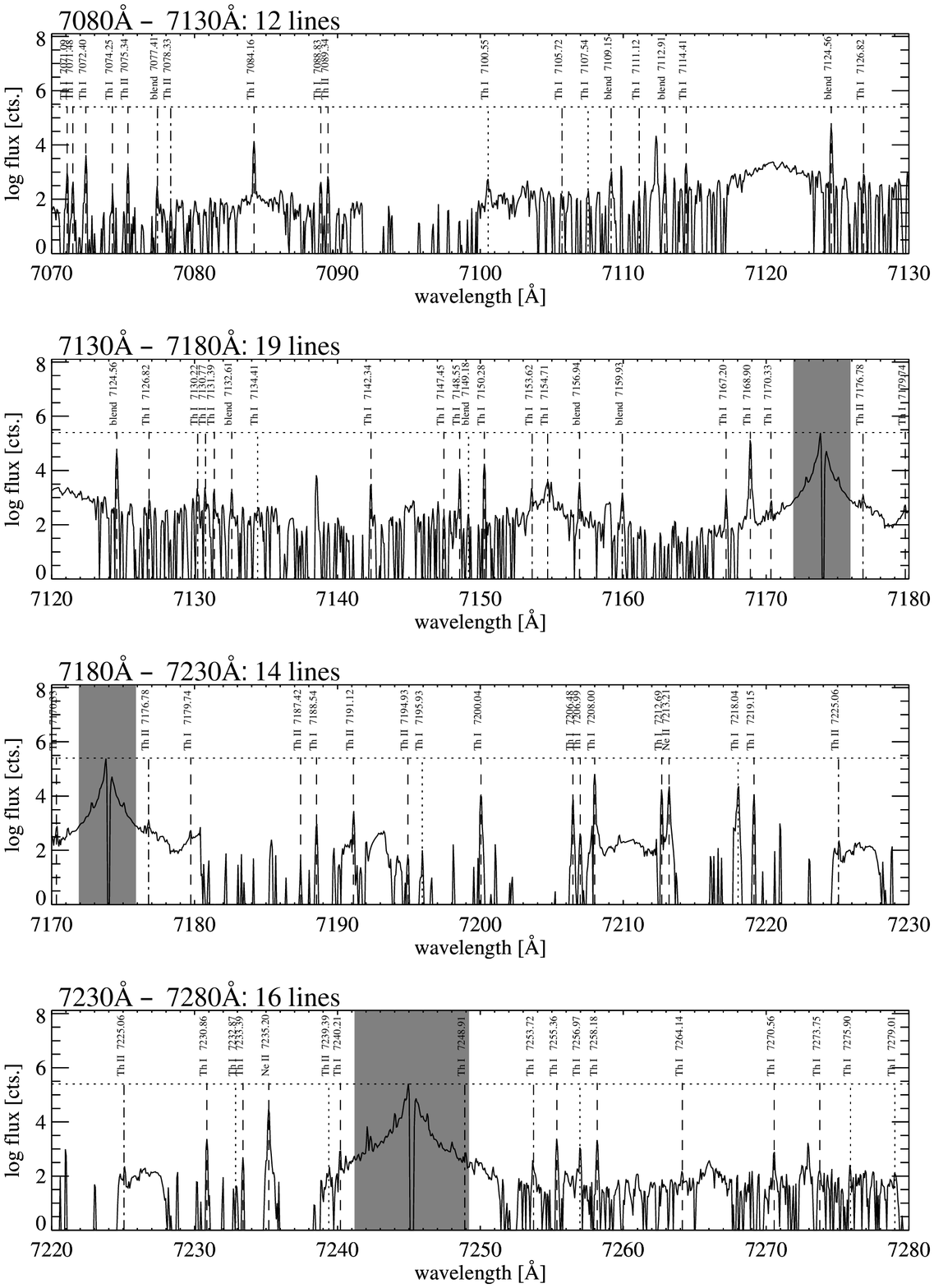}
\end{figure}
\clearpage
\begin{figure}
\includegraphics[width=\atlaswidth]{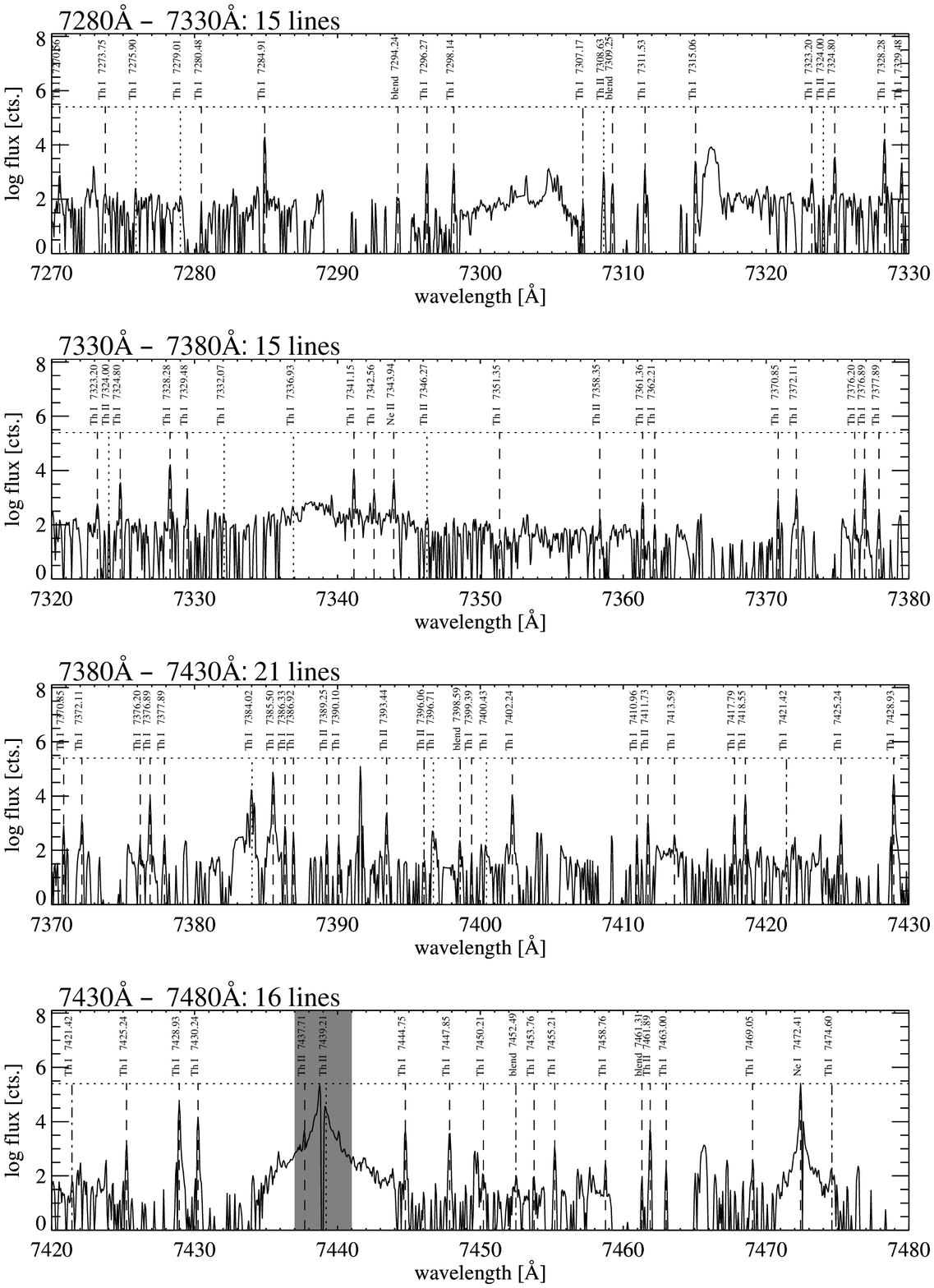}
\end{figure}
\clearpage
\begin{figure}
\includegraphics[width=\atlaswidth]{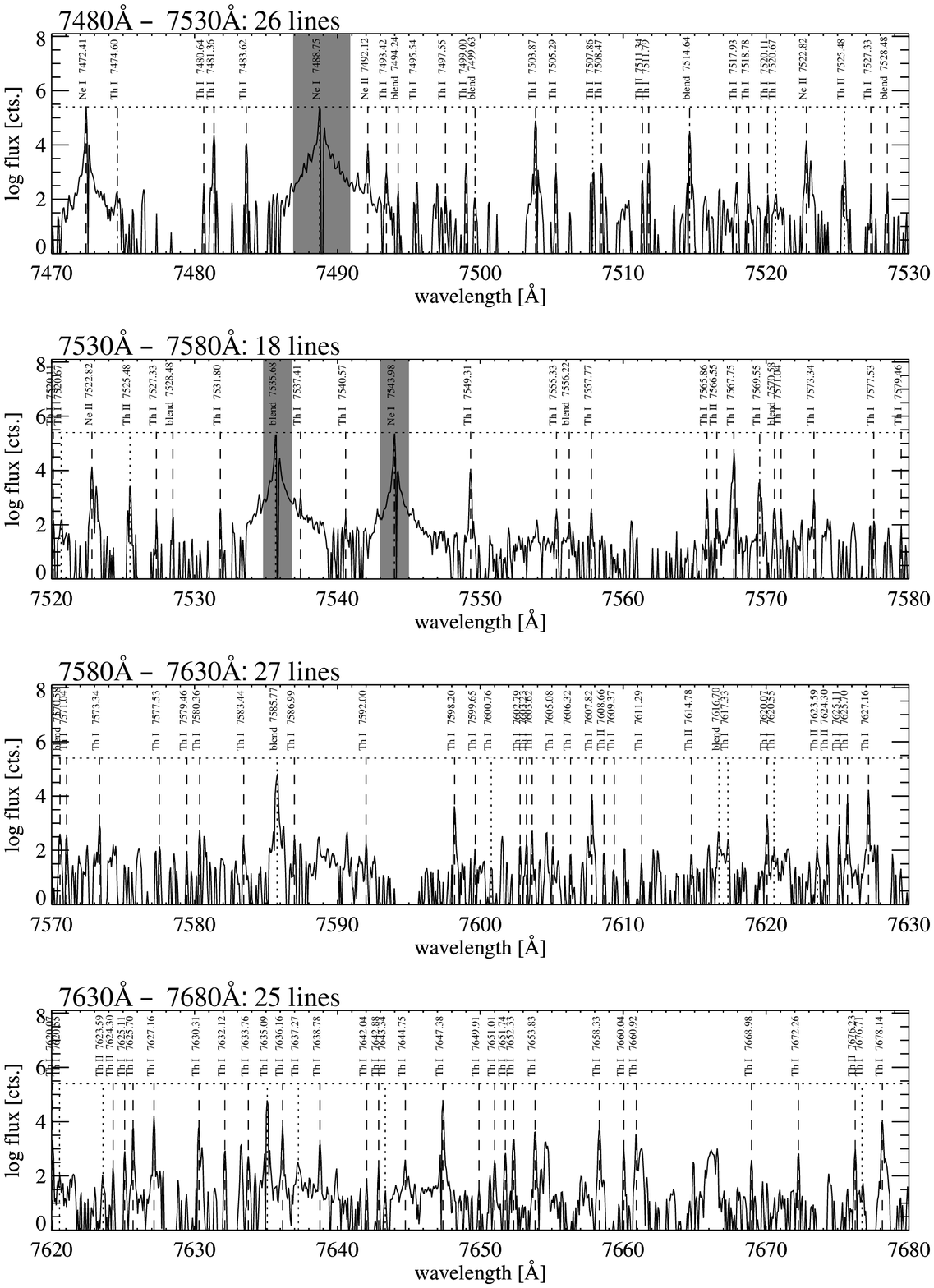}
\end{figure}
\clearpage
\begin{figure}
\includegraphics[width=\atlaswidth]{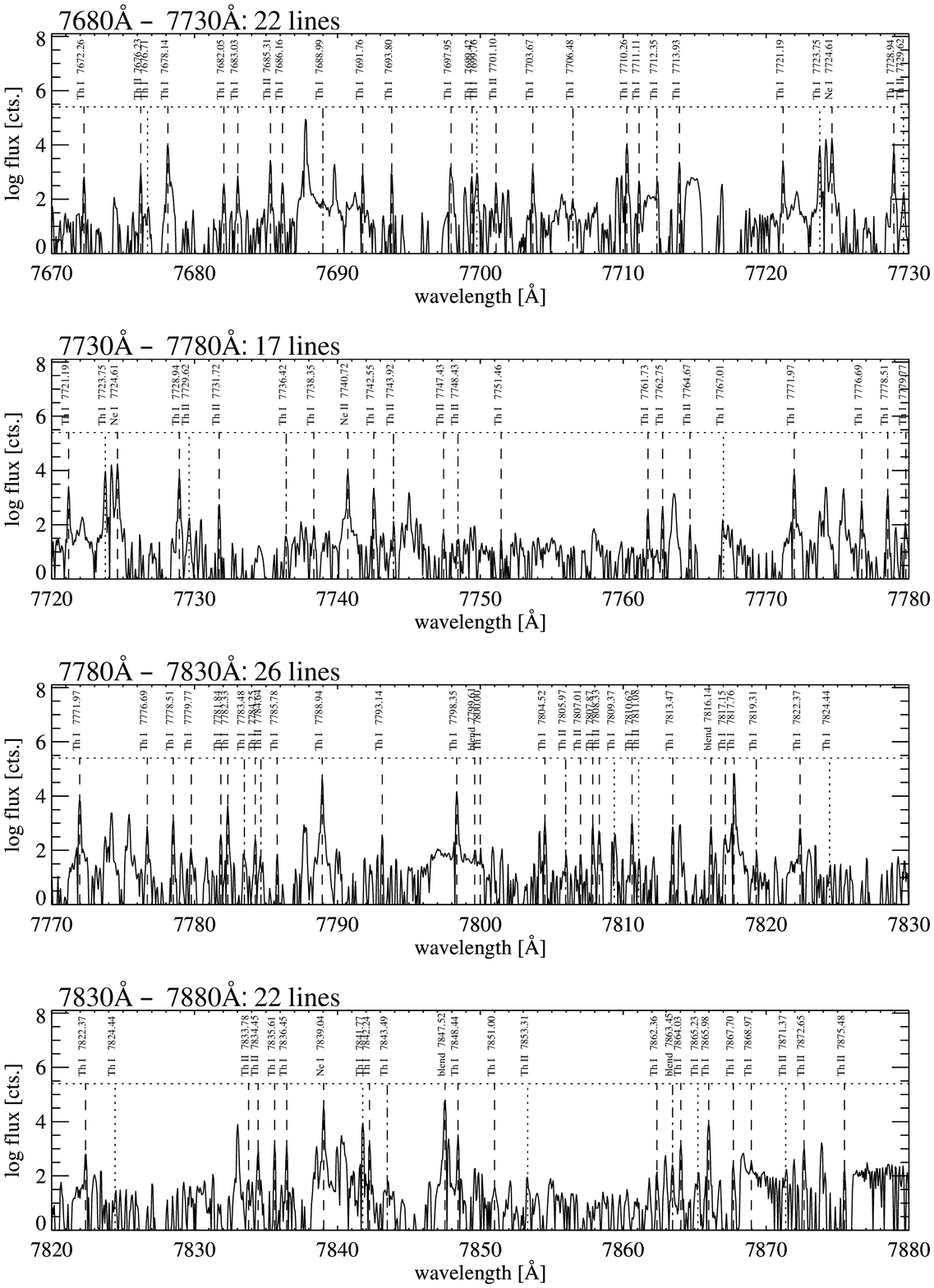}
\end{figure}
\clearpage
\begin{figure}
\includegraphics[width=\atlaswidth]{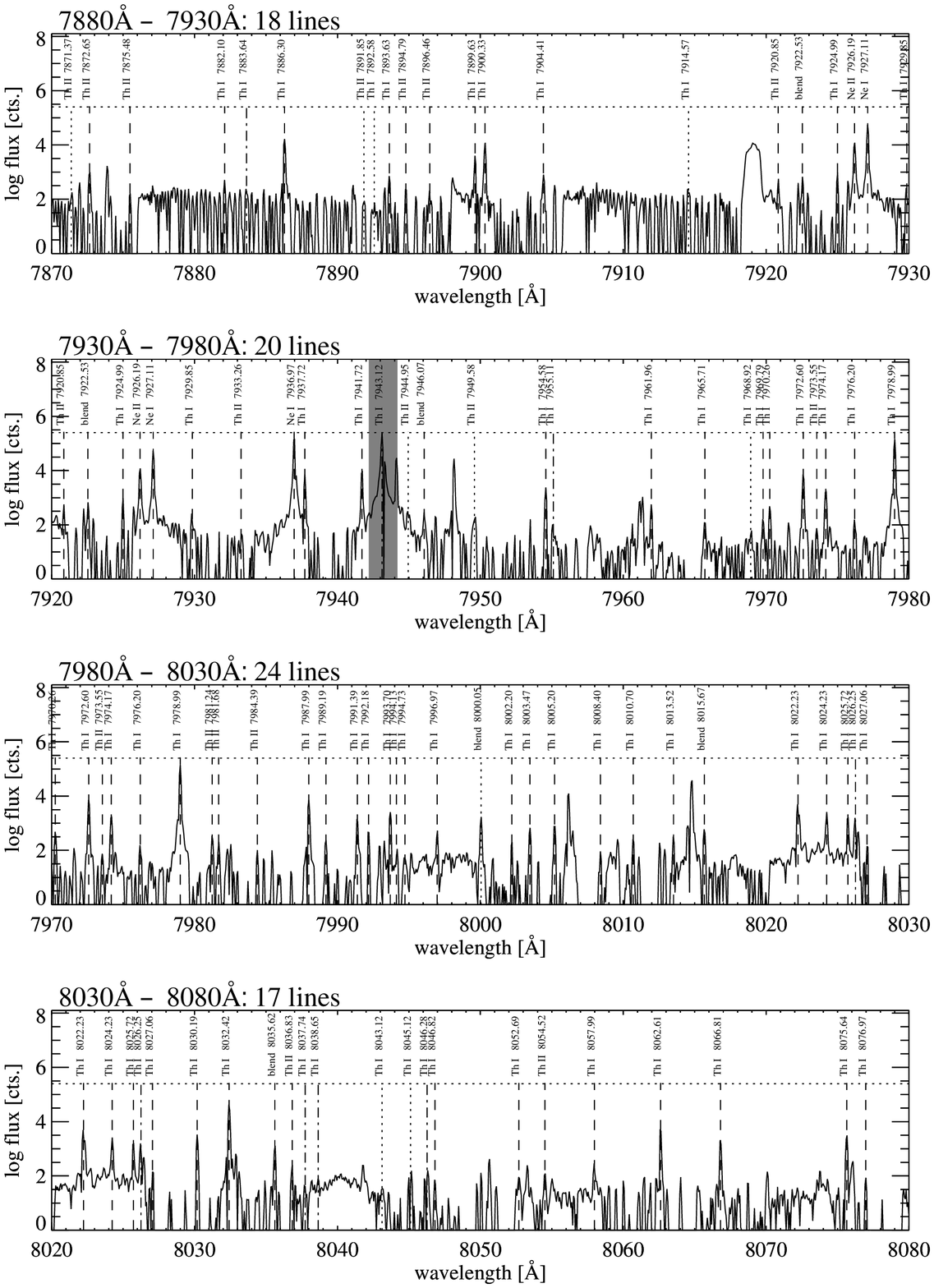}
\end{figure}
\clearpage
\begin{figure}
\includegraphics[width=\atlaswidth]{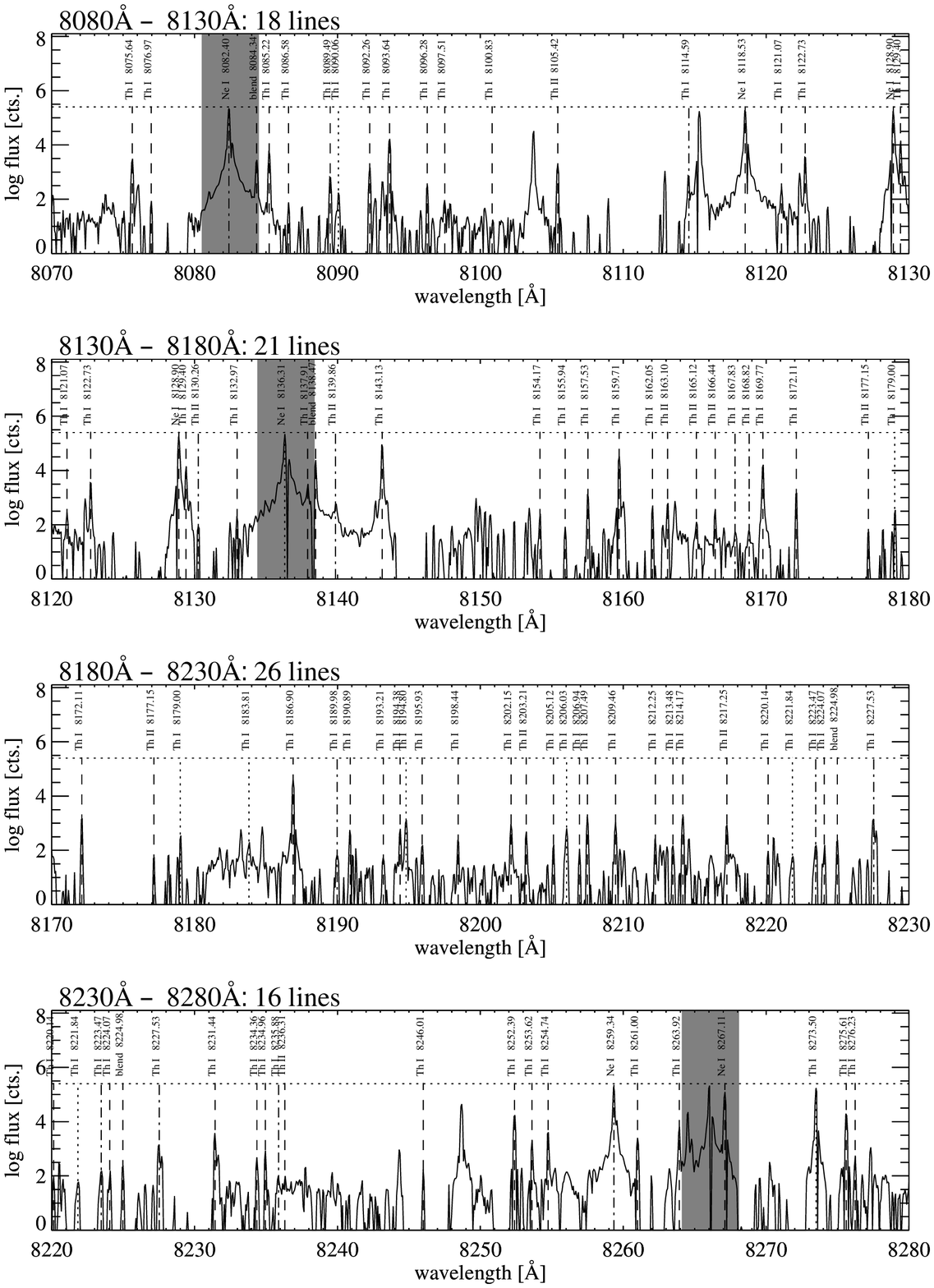}
\end{figure}
\clearpage
\begin{figure}
\includegraphics[width=\atlaswidth]{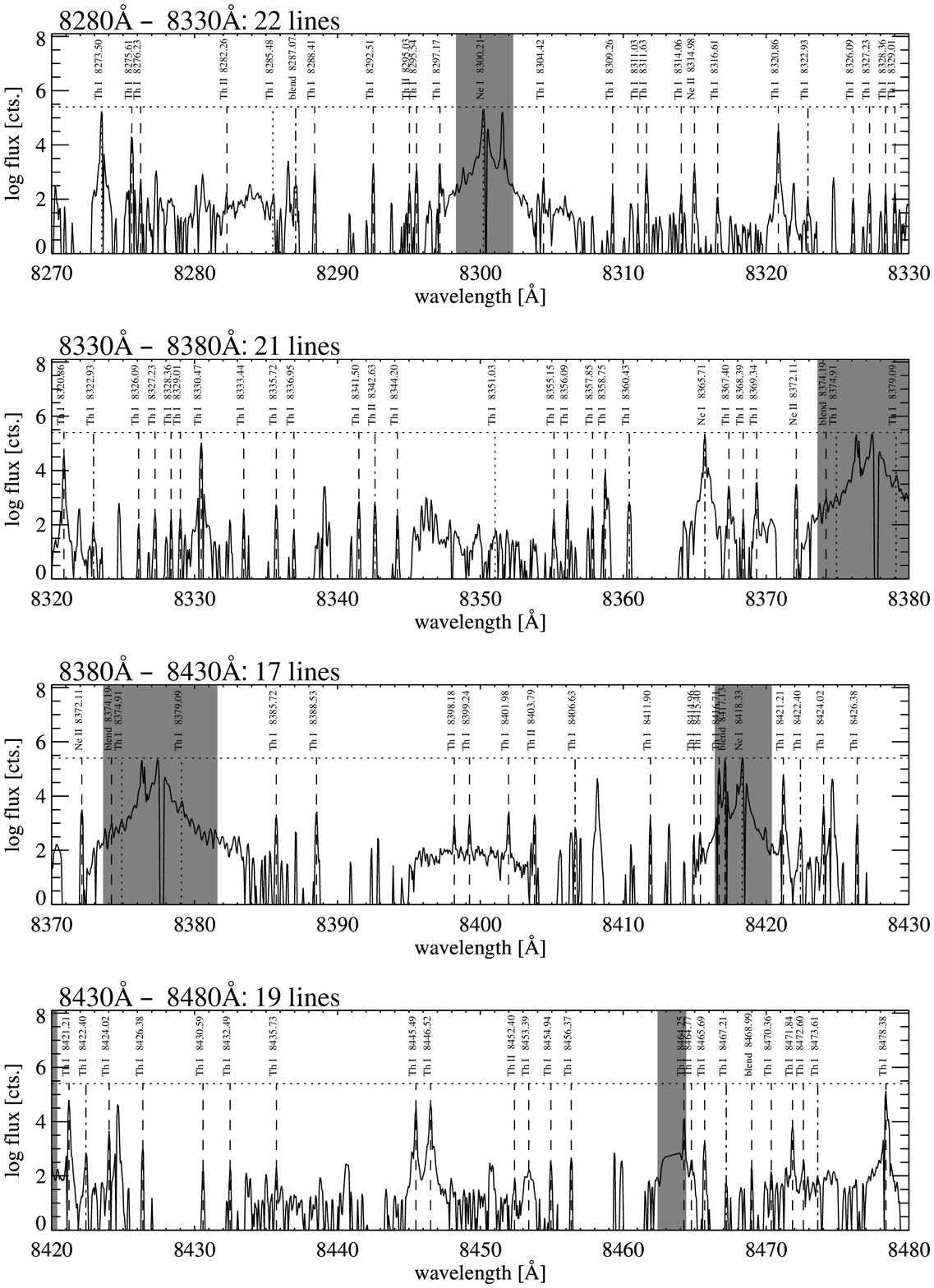}
\end{figure}
\clearpage
\begin{figure}
\includegraphics[width=\atlaswidth]{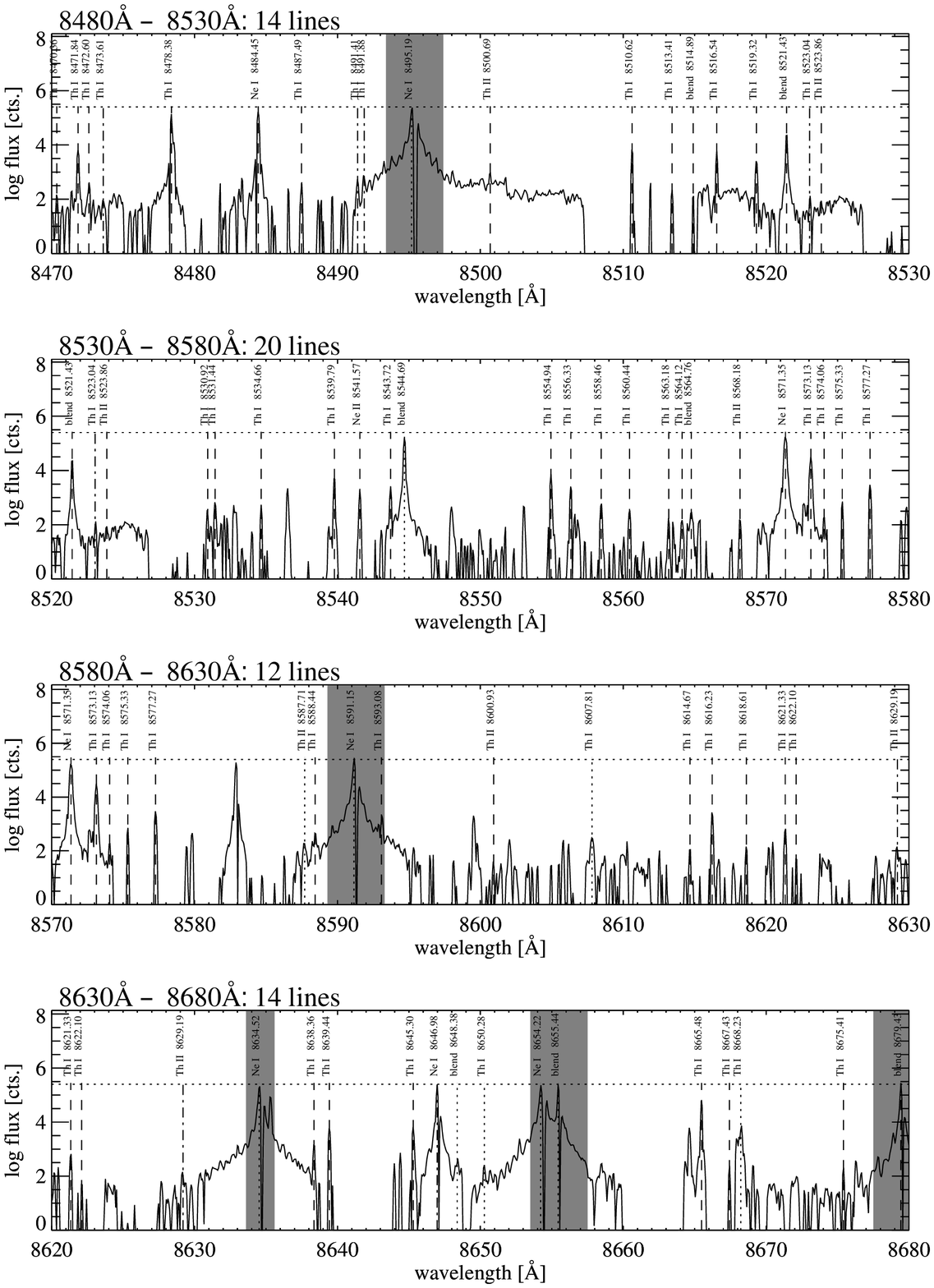}
\end{figure}
\clearpage
\begin{figure}
\includegraphics[width=\atlaswidth]{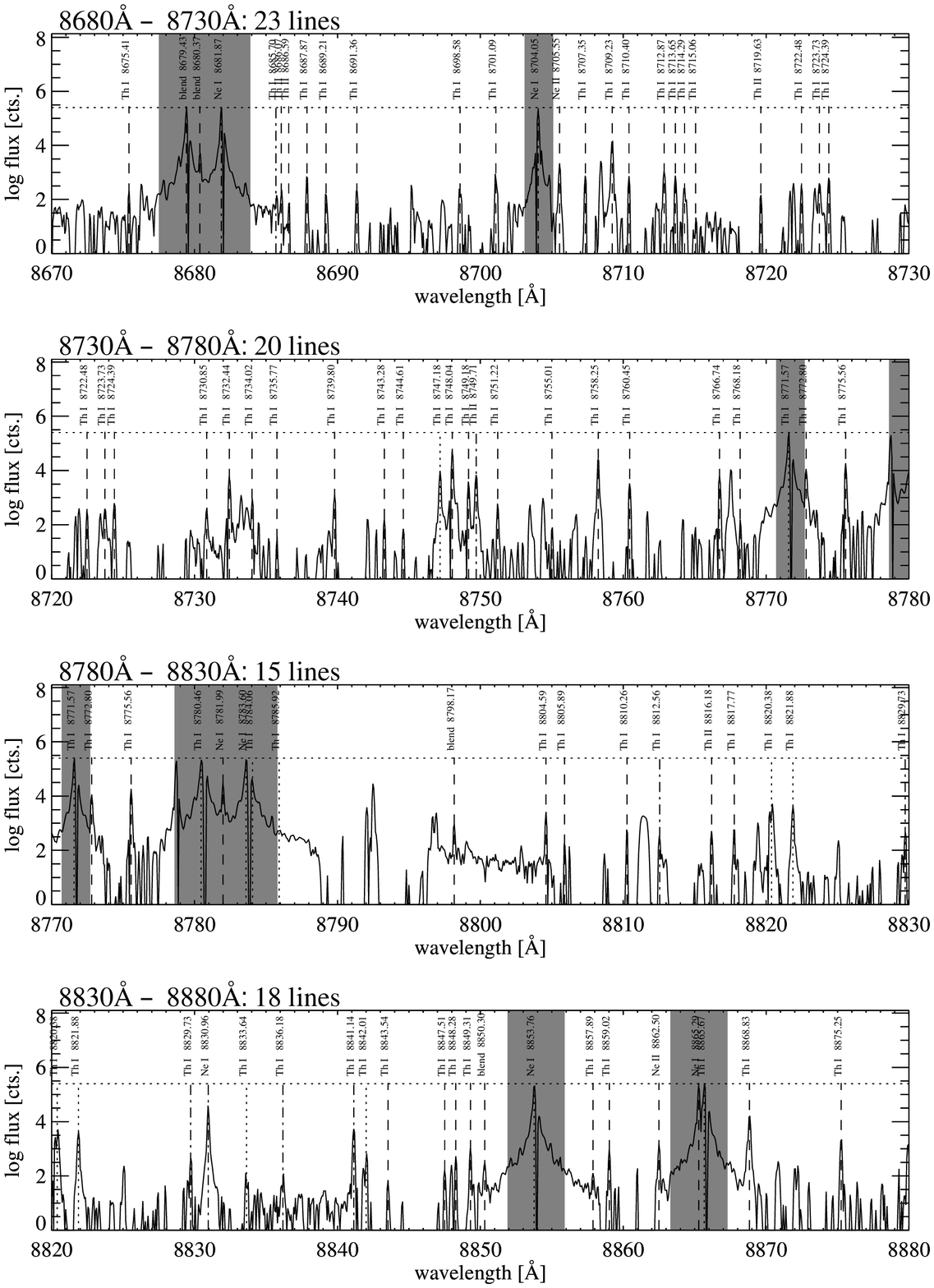}
\end{figure}
\clearpage
\begin{figure}
\includegraphics[width=\atlaswidth]{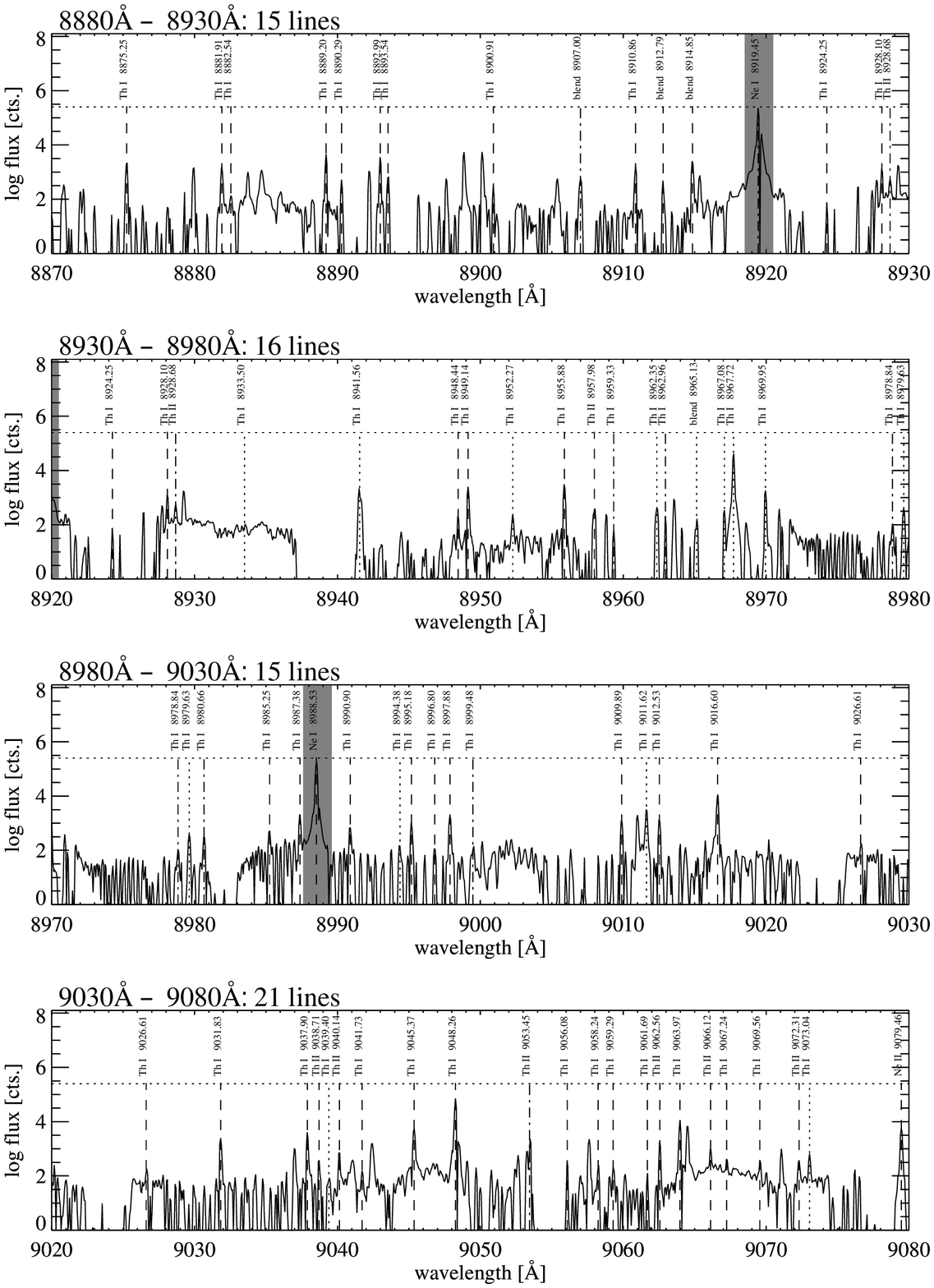}
\end{figure}
\clearpage
\begin{figure}
\includegraphics[width=\atlaswidth]{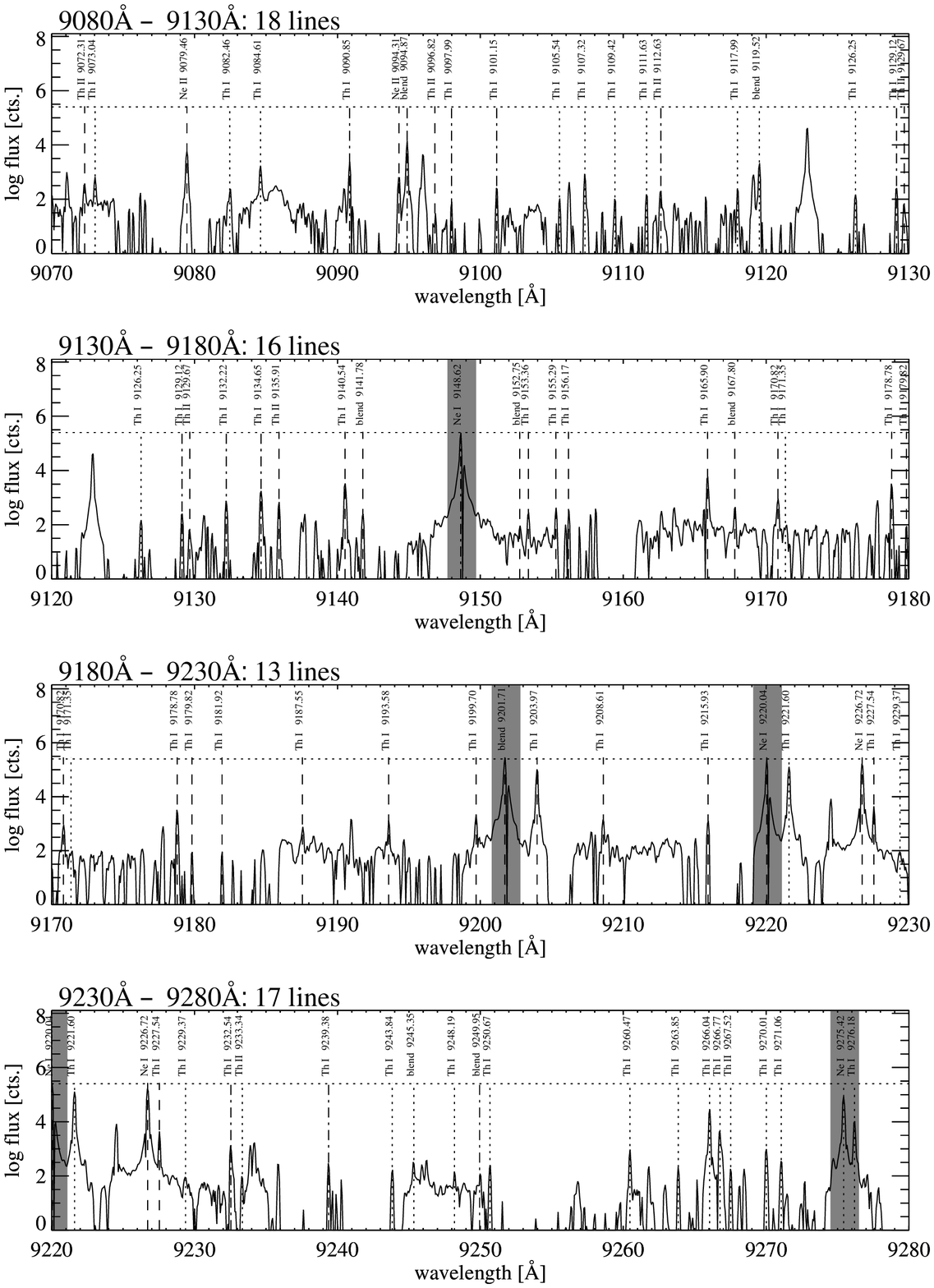}
\end{figure}
\clearpage
\begin{figure}
\includegraphics[width=\atlaswidth]{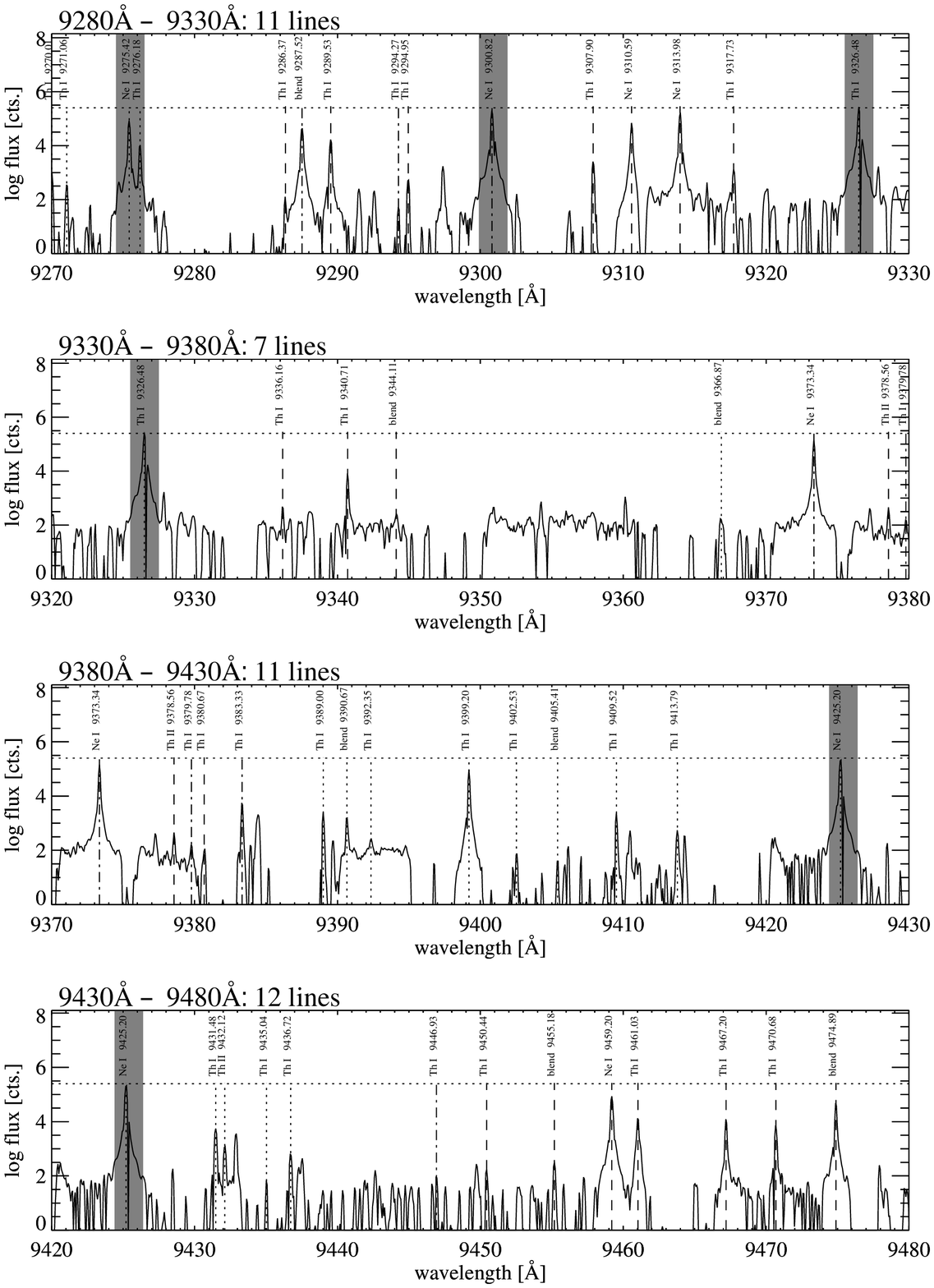}
\end{figure}
\clearpage
\begin{figure}
\includegraphics[width=\atlaswidth]{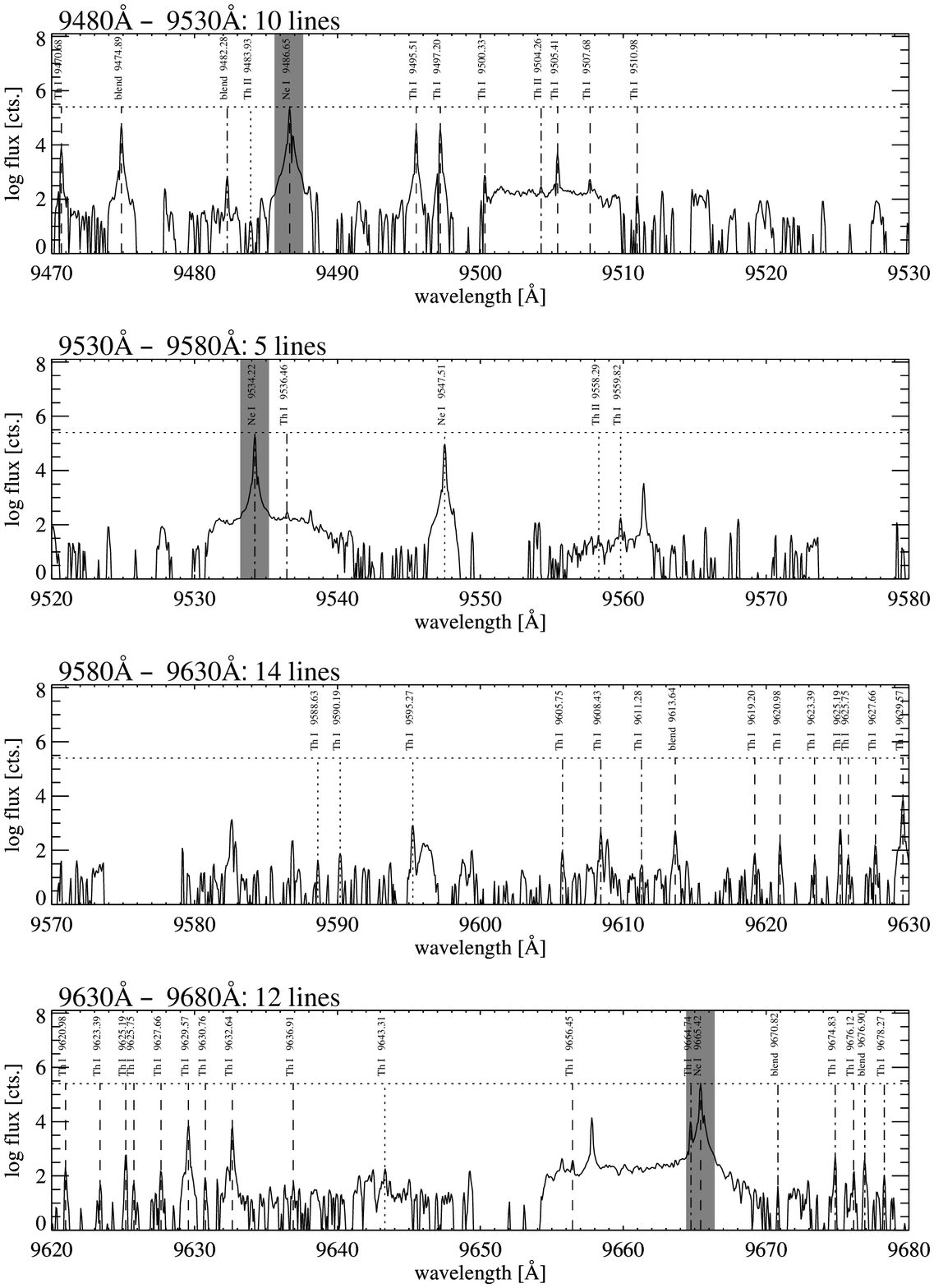}
\end{figure}
\clearpage
\begin{figure}
\includegraphics[width=\atlaswidth]{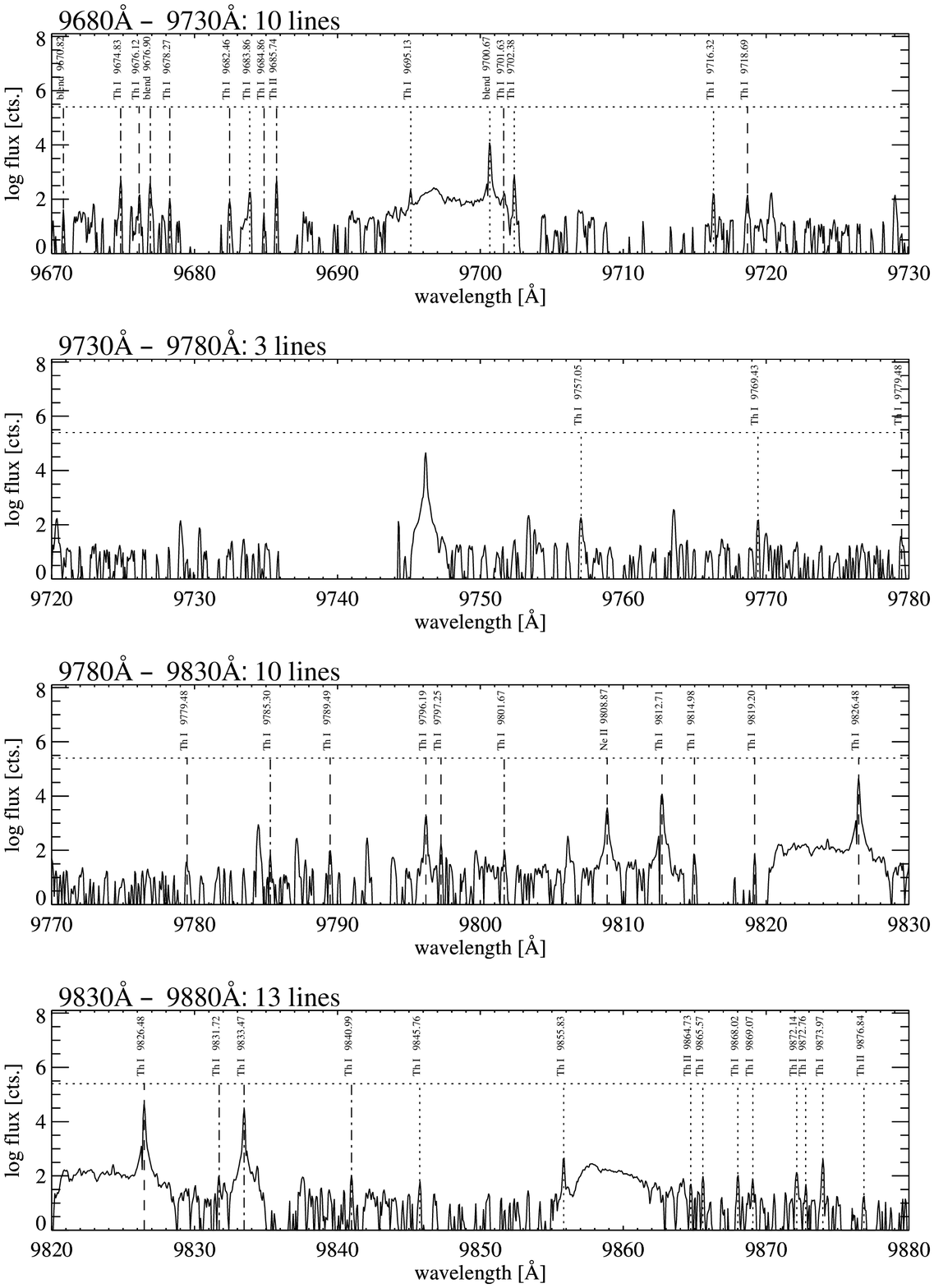}
\end{figure}
\clearpage
\begin{figure}
\includegraphics[width=\atlaswidth]{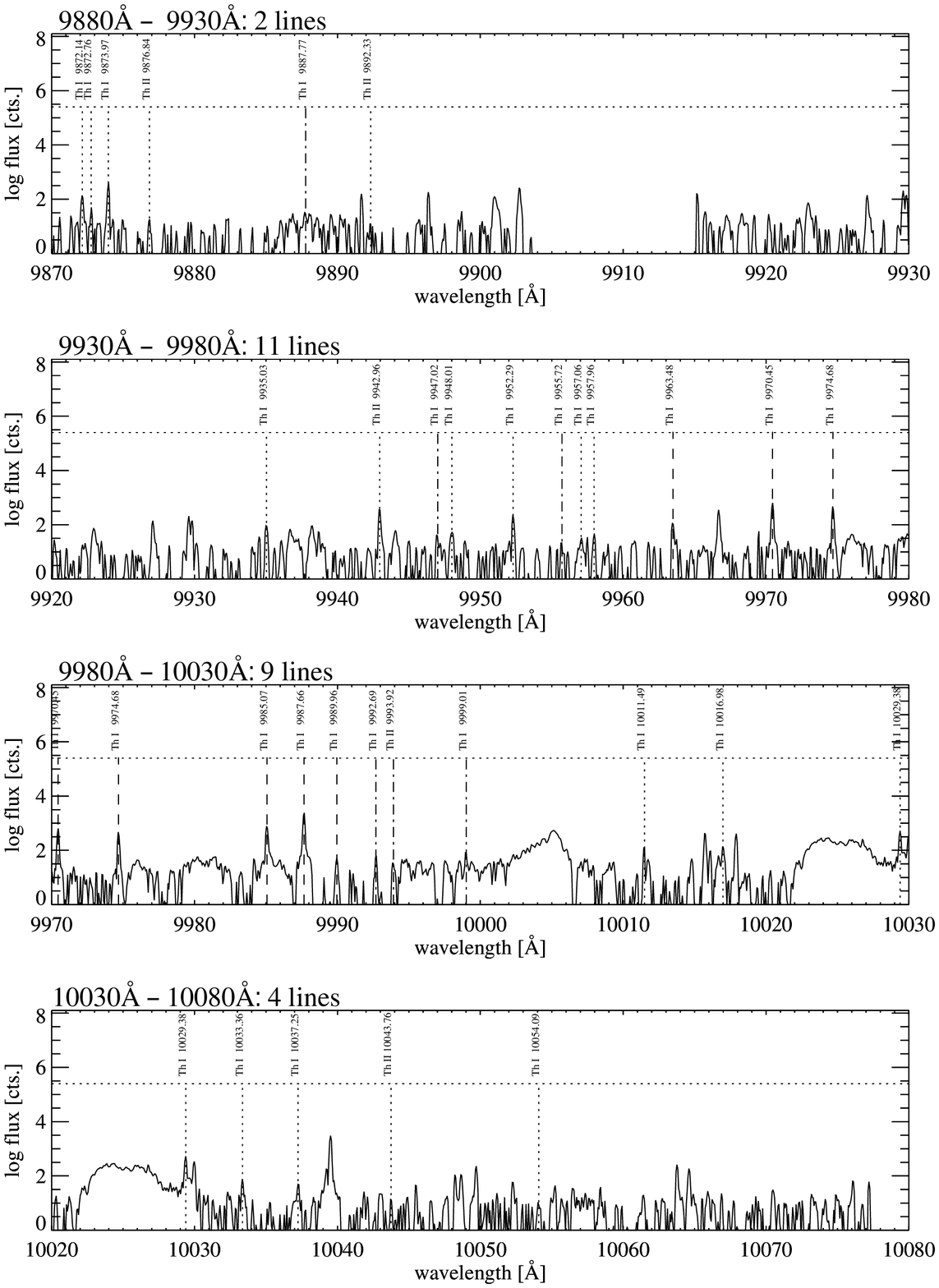}
\end{figure}
\clearpage
\begin{figure}
\includegraphics[width=\atlaswidth]{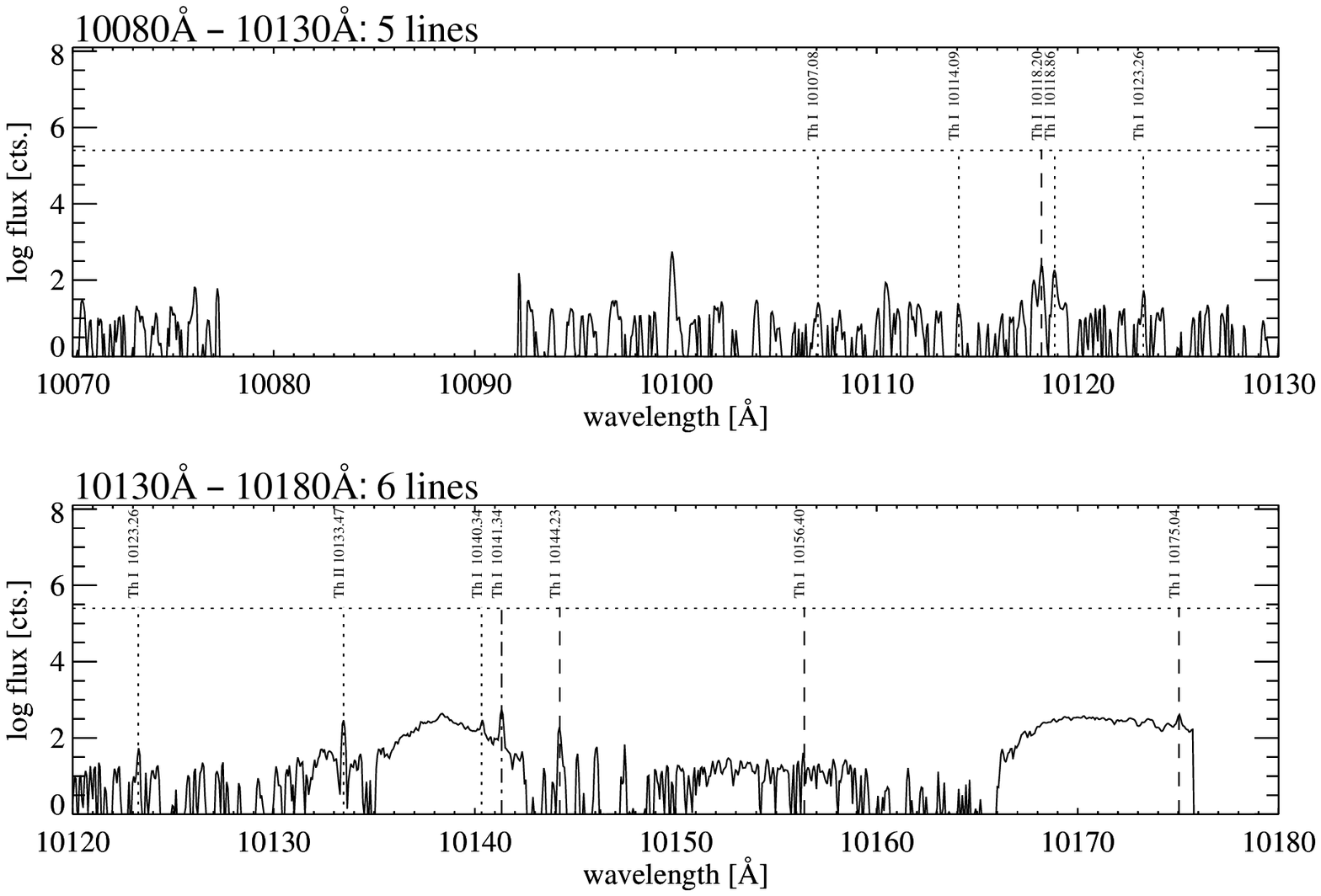}
\end{figure}
\clearpage

\end{document}